\def\Tr{\mathop{\rm Tr}\nolimits} %
\def\Sp{\mathop{\rm Sp}\nolimits} %
\def\bra#1{\langle #1 \,\vert}
\def\ket#1{\vert\, #1 \rangle}
\def\frac#1#2{{#1 \over #2}}
\def\Dirac#1{#1\hskip-6pt/}
\def\dd{\Dirac\partial}
\def\d{\mbox{d}}
\def\e{\mbox{e}}
\documentstyle[prd,aps]{revtex}
\begin{document}
\title{Electromagnetic form factors of the nucleon in the
chiral quark soliton model}
\author{Chr.V.~Christov\thanks{Permanent address:
Institute for Nuclear Research and Nuclear Energy, Sofia,
Bulgaria}$^{a,}$\thanks{christov@neutron.tp2.ruhr-uni-bochum.de},
A.Z.~G\'orski$^{b,}$\thanks{gorski@bron.ifj.edu.pl},
K.~Goeke$^{a,}$\thanks{goeke@hadron.tp2.ruhr-uni-bochum.de} and
P.V.~Pobylitsa$^{c,}$\thanks{maxpol@lnpi.spb.su}}
\address{$^a$Institut f\"ur Theoretische Physik II,
Ruhr-Universit\"at Bochum, D-44780 Bochum, Germany}
\address{$^b$Institute of Nuclear Physics, Radzikowskiego 152, 31-342
Cracow, Poland}
\address{$^c$Petersburg Nuclear Physics Institute, Gatchina,
St.Petersburg 188350, Russia}
\maketitle

\begin{abstract}
In this paper we present the derivation as well
as the numerical results for the electromagnetic form factors
of the nucleon within the chiral quark soliton model in the  semiclassical
quantization scheme. The model is based on semibosonized SU(2) Nambu --
Jona-Lasinio lagrangean, where the boson fields are treated as
classical ones. Other observables, namely the nucleon
mean squared radii, the magnetic moments, and the nucleon--$\Delta$
splitting are calculated as well. The calculations have been done taking into
account the quark sea polarization effects.
The final results, including rotational $1/N_c$ corrections, are
compared with the existing experimental data, and they are found to be
in a good agreement for the constituent quark mass of about 420 MeV.
The only exception is the neutron electric form factor which is overestimated.
\end{abstract}
\draft
\newpage

\section{Introduction}

In the last years one of the most challenging problems in elementary
particle physics seems to be the solution of QCD in the low energy region.
The main difficulties are due to the non-perturbative effects caused by the
growing effective coupling constant of the fundamental theory
in the low energy limit. This prevents one from using the well-known
main tool of theoretical physics --- the perturbation theory.
Because of this, the most intriguing features of QCD, confinement and
chiral symmetry breaking, still remain conceptual and practical
problems.

The above mentioned obstacles have initiated an increasing interest
in non-perturbative methods and effective
low-energy models of hadrons. The effective models are expected to
mimic the behavior of QCD at low energies.

The simplest purely fermionic Lorentz invariant model
describing spontaneous chiral symmetry breakdown is the Nambu--Jona-Lasinio
(NJL) model~\cite{Nambu61}. The NJL lagrangean
contains chirally invariant local four-fermion
interaction. In its simplest SU(2) form it has the
following structure:
\begin{equation}
{\cal L}_{NJL} = \bar \Psi \, (i \dd - m_0)\, \Psi +
{G\over 2} \, [(\bar \Psi \Psi )^2 + (\bar \Psi i \vec\tau
\gamma_{5} \Psi )^2\,] \ ,
\label{NJLLAG} \end{equation}
where $\Psi$ is the quark field, $G$ is the coupling constant,
$\vec\tau$ are the Pauli matrices in the isospin space and
$m_0$ is the current mass the same for both $up$ and $down$
quarks.

Applying the well known
bosonization procedure~\cite{Eguchi74}
the NJL model is expressed in terms of auxiliary meson fields
$\sigma, {\vec\pi}$:
\begin{equation}
\bar \Psi \  [ \ i \dd - m_0-
g (\sigma + i\, {\vec\pi\cdot\vec\tau}\, \gamma_5 )\ ] \, \Psi -
{g^2\over {2G}} (\sigma^2 + {\vec\pi}^2).
\label{NEWLAG} \end{equation}
Here, $g$ is the physical
pion--quark coupling constant implying that $\vec\pi$
are the physical pion fields.
We assume that the meson fields are constrained to the chiral circle:
\begin{equation}
\sigma^2({\vec x}) + {\vec\pi}^2({\vec x}) = f^2_{\pi} \ ,
\label{CHIRCIR}
\end{equation}
where $f_{\pi}=93$ MeV is the pion decay constant.

In fact, an equivalent effective chiral quark meson theory~\cite{Diakonov84}
can be derived from the instanton model of the QCD vacuum.

In the chiral quark soliton model~\cite{Diakonov89,Reinhardt88,TMeissner89}
based on the lagrangean (\ref{NEWLAG}) (frequently referred to
simply as NJL
model) the baryons appear as a bound state of $N_c$ valence quarks,
coupled to the polarized Dirac sea. Actually, the
baryon solution of the model is obtained in two steps. In the first step,
motivated by the large $N_c$ limit, a static localized solution
(soliton) is found  by solving the corresponding equations of motion
in an iterative self-consistent procedure~\cite{Reinhardt88,TMeissner89},
assuming that the $\sigma$ and $\vec\pi$ fields have hedgehog structure.
However, this hedgehog soliton does not preserve the spin and isospin. In
order to describe the nucleon properties one needs nucleon states with
proper spin and isospin numbers. To this end, making use of the rotational
zero modes, the soliton is quantized~\cite{Diakonov89,Reinhardt89}.
Within the semiclassical quantization
scheme quite successful calculations have been reported for the nucleon-delta
mass splitting~\cite{Goeke91} and axial-vector form
factor~\cite{TMeissner91} as well as some results for the
nucleon electric form factors~\cite{Gorski92}. Very recently, in the
semiclassical quantization procedure important $1/N_c$ rotational
corrections have been derived~\cite{Christov94} for
isovector magnetic moment as well as for the axial coupling constant.
These corrections are derived in a formalism~\cite{Christov94}, based on the
path integral approach~\cite{Diakonov89}. The scheme involves a
time-ordering of collective operators which follows from the collective path
integral. It improves considerably the theoretical values for the isovector
magnetic moment, which is strongly underestimated~\cite{Wakamatsu91} in the
leading order.

Besides its numerical success, there are also theoretical reasons why the
NJL  model, described by lagrangean (\ref{NEWLAG}), is considered as the one
of the most promising effective theories describing low energy QCD phenomena.

First, the model is the simplest quark model
describing  spontaneous breaking of the chiral symmetry, a basic
feature of QCD. This allows to treat the pions as Goldstone particles
and to include familiar methods of the current algebra
into the model. As for the confinement, which is absent in the model,
some recent lattice QCD results for the vacuum correlation functions
of hadronic currents~\cite{Chu94} suggest that the confinement mechanism
seems to have little impact on the light hadron structure, in particular on
the nucleon and Delta structure.

Second, there are various hints~\cite{Diakonov84,Ball89,Schaden90}  that the
NJL-type lagrangean can be obtained from QCD in various low-energy
approximations. It should be stressed that the large $N_c$ limit plays a
prominent role in those considerations.

The aim of this paper is to calculate the electromagnetic nucleon form
factors within the $SU(2)$ chiral quark soliton model, based on the
semibosonized NJL-type lagrangean, where the vacuum polarization effects
taken into account. The calculations done in the semiclassical quantization
scheme will include the rotational $1/N_c$ corrections,
which have been shown~\cite{Christov94} to be important for the
isovector magnetic moment. In the derivation of the matrix element of
the electromagnetic current in the semiclassical quantization scheme
not-commuting collective operators appear and the final result depends on
the ordering of these operators. In fact, similar to ref.~\cite{Christov94}
the problem of the proper ordering of the collective operators is solved
in the present paper using a path-integral approach~\cite{Diakonov89} to the
semiclassical quantization, which allows a natural access to
the problem. This scheme allows to calculate the nucleon electromagnetic
form factors, as well as
such quantities like the electric mean squared radii, the magnetic moments,
the nucleon--$\Delta$ energy splitting, and the electric and magnetic
charge distributions of the nucleon.

The paper is organized as follows. In Section 2 we derive general
expressions for the nucleon form factors in the NJL model. They are
split in Dirac sea and valence quark contributions both including
rotational corrections, coming from the expansion of the effective action
in powers of the angular velocity of the soliton.
Since the NJL model is non-renormalizable an ultraviolet
regularization, which preserves the electromagnetic gauge invariance,
is introduced in Section 3. The general formalism developed in Sections 2
and 3 is then applied to evaluate the nucleon electromagnetic form factors
in Section 4. In Section 5 we present and discuss our numerical results.

\section{Dirac sea and valence quark contributions to
nucleon form factors}

Our main goal in this section is to derive a general formula
for nucleon form factors in the NJL model including rotational corrections.
We start with the path integral representation for the nucleon
matrix element of the corresponding quark current.
After integrating out the quark fields we find that the result
splits in two parts, the Dirac sea and the valence quark
contributions. Usually the remaining integral over
the meson fields is evaluated
within the saddle point approximation.
However, the saddle-point classical solution has a hedgehog symmetry
that breaks both rotational
and $SU(2)$ isospin symmetries. In order to restore them
one has to consider a rotating soliton and to perform a functional
integration over the time-dependent orientation matrix of the soliton.
Expanding in powers of the angular velocity
up to linear order we derive rotational corrections
to the nucleon form factors. A special attention is paid to the correct
ordering of non-commuting collective operators describing the rotational
degrees of freedom of the soliton. In this section the consideration
is limited to the non-regularized case and the ultraviolet
regularization is postponed till Section 3.

Under the chiral circle restriction (\ref{CHIRCIR})
we can introduce the $SU(2)$ chiral meson field
\begin{equation}
U=\frac{1}{f_\pi}(\sigma + i\, {\vec\pi\cdot\vec\tau})
\end{equation}
and work with the Euclidean lagrangean
\begin{equation}
L=\Psi^\dagger D(U) \Psi\,,
\label{action-quark-pion}
\end{equation}
where the operator
\begin{equation}
D(U) = \gamma_4 (\gamma_\mu \partial_\mu + M U^{\gamma_5} + m_0)
=  \partial_\tau + h(U)
\label{D-U-definition}
\end{equation}
includes the Euclidean time derivative $\partial_\tau$ ($\tau=x_4$) and
the Dirac one-particle hamiltonian reads:
\begin{equation}
h(U)=  \gamma_4 (\gamma_k \partial_k + M U^{\gamma_5} + m_0).
\label{HAMIL}
\end{equation}
Here $M=gf_{\pi}$ is the constituent quark mass. For convenience, we choose
to work with hermitean Euclidean Dirac matrices $\gamma_\alpha^\dagger =
\gamma_\alpha$. We use notation
\begin{equation}
U^{\gamma_5} = \frac{1+\gamma_5}{2} U + \frac{1-\gamma_5}{2} U^\dagger.
\end{equation}

The aim of this paper is to calculate the electromagnetic
nucleon form factors. However, to be more general we start with a matrix
element of quark current
$\Psi^\dagger O \Psi$, where $O$ is some matrix with spin and isospin
indices, and represent this matrix element
by the Euclidean functional integral~\cite{Diakonov89}
with the lagrangean (\ref{action-quark-pion}):
\begin{eqnarray}
&&\langle N^\prime, {\vec p}^\prime | \Psi^\dagger(0) O \Psi(0) | N,
{\vec p} \rangle
=\lim_{T \to+\infty } \frac 1Z \int \d^3 x\, \d^3y\,
\e^{- i{\vec p}^\prime {\vec x}^\prime +
i{\vec p} {\vec x}}\nonumber \\
&&\int{\cal D}U \int {\cal D}\Psi \int {\cal D}\Psi^\dagger
J_{N^\prime}(\vec x^\prime,-T/2) \> \Psi^\dagger(0) O \Psi(0) \>
J_N^\dagger(\vec x, T/2) \e^{- \int d^4 z  \Psi^\dagger D(U) \Psi }\,.
\label{form-factor-integral}
\end{eqnarray}
Here the nucleon state is created by the nucleon current $J_N^\dagger(x)$
constructed of $N_c$ (number of colors) quark fields $\Psi$
\begin{equation}
J_N(x)=\frac
1{N_c!}\epsilon^{c_1\ldots c_{N_c}}\Gamma^{\{\alpha_1\ldots
\alpha_{N_c}\}}_{JJ_3,TT_3}
\Psi_{ c_1\alpha_1}(x)\cdots\Psi_{ c_{N_c}\alpha_{N_c}}(x)
\label{baryon-current}
\end{equation}
where $ c_i$ are color indices and $\Gamma^{\alpha_1\cdots \alpha_{N_c}}_{
JJ_3,TT_3}$ is a matrix with $\alpha_i$ standing for both flavor and
spin indices. $J$ and $T$ denote the nucleon spin and isospin,
respectively. In the limit of large Euclidean times $x_4^\prime,x_4$ only
the lowest nucleon state survives in (\ref{form-factor-integral}).

We assume that the constant $Z$ in eq.~(\ref{form-factor-integral}) is
chosen so that the nucleon states obey the non-relativistic normalization
condition:
\begin{equation}
\langle N, {\vec p}^\prime |  N, {\vec p} \rangle
= (2\pi)^3 \delta^{(3)}({\vec p}^\prime -{\vec p}) \, .
\end{equation}

Let us integrate the quarks out in (\ref{form-factor-integral}).
The result is naturally split in valence and Dirac sea parts (see
Fig.\ref{Figr1}):
\begin{equation}
\langle N^\prime, {\vec p}^\prime | \Psi^\dagger O \Psi | N, {\vec p} \rangle
= \langle N^\prime, {\vec p}^\prime | \Psi^\dagger O \Psi | N, {\vec p}
\rangle^{sea}
+ \langle N^\prime, {\vec p}^\prime | \Psi^\dagger O \Psi | N, {\vec p}
\rangle^{val}\,,
\end{equation}
where
\begin{eqnarray}
&&  \langle N^\prime, {\vec p}^\prime | \Psi^\dagger O \Psi | N, {\vec p}
\rangle^{sea}
= N_c \lim_{T\to+\infty} \frac 1Z \int \d^3 x\, \d^3y\,
\e^{- i{\vec p}^\prime {\vec x}^\prime + i{\vec p}
{\vec x}} \nonumber \\
&&
\times \int{\cal D}U\,
\Gamma_{N^\prime}^{\beta_1 \ldots \beta_{N_c}}
\Gamma_N^{\alpha_1 \ldots \alpha_{N_c}\ast}
\prod\limits_{k=1}^{N_c}
\langle x^\prime,T/2| \frac{1}{D(U)}
|x,-T/2\rangle_{\beta_k\alpha_k}
\Sp\Bigl\{O_{\gamma\gamma^\prime} \langle 0,0 |
\frac{-1}{D(U)} | 0,0\rangle_{\gamma^\prime\gamma}\Bigr\}\nonumber\\
&&\times
\e^{N_c\Tr\log [D(U)]} \,
\label{sea-contribution}
\end{eqnarray}
and
\begin{eqnarray}
&&\langle N^\prime, {\vec p}^\prime | \Psi^\dagger O \Psi(0) | N, {\vec p}
\rangle^{val}
=N_c \lim_{T\to+\infty}\frac 1Z \int \d^3 x \d^3y
\e^{- i{\vec p}^\prime {\vec x}^\prime +
i{\vec p} {\vec x}} \nonumber \\
&&\times
 \int {\cal D}U \, \Gamma_{N^\prime}^{\beta_1 \ldots \beta_{N_c}}
\Gamma_N^{\alpha_1 \ldots \alpha_{N_c}\ast}
\prod\limits_{k=2}^{N_c}
\langle \vec x^\prime,T/2 | \frac{1}{D(U)} |\vec x,-T/2
\rangle_{\beta_k\alpha_k} \nonumber \\
&& \times
\langle x^\prime,T/2 | \frac{1}{D(U)} |  0,0 \rangle_{\beta_1\gamma}
O_{\gamma\gamma^\prime}\langle  0,0|  \frac{1}{D(U)} | \vec x, -T/2
\rangle_{\gamma^\prime\alpha_1} \e^{N_c\Tr\log [D(U)]} \,.
\label{valence-contribution} \end{eqnarray}

In the limit of large number of colors $N_c$
the functional integrals over the chiral field $U(x)$
in eqs.~(\ref{sea-contribution}), (\ref{valence-contribution})
can be evaluated by the saddle point method.
The saddle point solution $\bar U$
is a static localized field configuration with the hedgehog symmetry
\begin{equation}
\bar U(x)= \e^{i P(|{\vec x}|) (x^a \tau^a)/|{\vec x}|}\,,
\end{equation}
where $P(|{\vec x}|)$ is the profile function of the soliton.
This saddle point solution breaks the space rotational and isospin
symmetries.
Therefore, even in the leading order in $N_c$, one should go beyond the
field $\bar U$, extending the path integral over all fields of the form:
\begin{equation}
U(x)= R(x_4) \bar U({\vec x}) R^\dagger(x_4)\,,
\label{rotating-soliton}
\end{equation}
where $R(x_4)$ is the time-dependent $SU(2)$ orientation matrix of the
soliton.

For this ansatz (\ref{rotating-soliton})
the operator $D(U)$ (\ref{D-U-definition})
can be written as
\begin{equation}
D(U) = R\, [D(\bar U)+ R^\dagger \dot R]\, R^\dagger \,.
\label{D-U-rotating}
\end{equation}
The dot stands for the derivative with respect to the Euclidean time $x_4$.
Similarly, the quark propagator in the background meson field $U$ can be
rewritten as
\begin{equation}
\bra{x}\,\frac 1{D(U)}\, \ket{x^\prime}\,=\,R(x_4)\,\bra{x}\,\frac 1{D(\bar
U)+R^\dagger \dot R}\, \ket{x^\prime}\,R^\dagger(x^\prime_4)\,.
\label{Eq10b}\end{equation}

Let us insert ansatz (\ref{rotating-soliton}) into our formulas for the
matrix element~(\ref{sea-contribution}), (\ref{valence-contribution}) and
replace the functional integral over $U(x)$
by the integral over the time dependent orientation matrix $R(x_4)$:
\begin{eqnarray}
&&  \langle N^\prime, {\vec p}^\prime | \Psi^\dagger O \Psi | N, {\vec
p} \rangle^{sea}
= N_c \lim_{T\to+\infty}\frac 1Z \int \d^3 x \d^3y
\e^{- i{\vec p}^\prime {\vec x}^\prime +
i{\vec p} {\vec x}} \nonumber \\
&&
\times
\int{\cal D}R \,
\Gamma_{N^\prime}^{\beta_1 \ldots \beta_{N_c}}
\Gamma_N^{\alpha_1 \ldots \alpha_{N_c}\ast}
\prod\limits_{k=1}^{N_c} \Bigl[R(T/2)
\langle \vec x^\prime,T/2| \frac{1}{D(U)} |\vec x,-T/2\rangle
R^\dagger (-T/2)\Bigr]_{\beta_k\alpha_k}\nonumber \\
&&\times\Sp\Bigl\{O_{\gamma\gamma^\prime} \Bigl[R(0)\langle 0,0 |
\frac{-1}{D(U)} |
0,0\rangle R^\dagger (0)\Bigr]_{\gamma^\prime\gamma}\bigr\}\e^{N_c\Tr\log
[D(\bar U)+ R^\dagger \dot R]} \label{sea-contributionR}
\end{eqnarray}
and
\begin{eqnarray}
&&\langle N^\prime, {\vec p}^\prime | \Psi^\dagger O \Psi(0) | N, {\vec
p} \rangle^{val}
=N_c \lim_{T\to+\infty}\frac 1Z \int \d^3 x \d^3y
\e^{- i{\vec p}^\prime {\vec x}^\prime +
i{\vec p} {\vec x}} \nonumber \\
&&\times
 \int {\cal D}R \,\e^{N_c\Tr\log [D(\bar U)+ R^\dagger \dot R]}
\Gamma_{N^\prime}^{\beta_1
\ldots \beta_{N_c}} \Gamma_N^{\alpha_1 \ldots \alpha_{N_c}\ast}
\prod\limits_{k=2}^{N_c}\Bigl[R(T/2)
\langle \vec x^\prime,T/2 | \frac{1}{D(U)} |\vec x,-T/2\rangle
R^\dagger (-T/2)\Bigr]_{\beta_k\alpha_k} \nonumber \\
&& \times
\Bigl[R(T/2)\langle \vec x^\prime,T/2 | \frac{1}{D(U)} |  0,0
\rangle R^\dagger (0)\Bigr]_{\beta_1\gamma}  O_{\gamma\gamma^\prime}
\Bigl[\langle  0,0|  \frac{1}{D(U)} | \vec x,-T/2\rangle R^\dagger
(-T/2)\Bigr]_{\gamma^\prime\alpha_1}\,. \label{valence-contributionR}
\end{eqnarray}

In eqs.~(\ref{sea-contributionR}), (\ref{valence-contributionR}) despite of
ansatz (\ref{rotating-soliton}) we still have a complicated path integral
over ${\cal D}R$. In order to solve this problem we make essentially use
that in the large $N_c$ limit the angular
velocity $R^\dagger \dot R$ of the soliton as well as its derivative is
suppressed~\cite{Diakonov89}. Hence, similarly to ref.~\cite{Diakonov89} we
can use expansions in powers of it
\begin{equation}
N_c \Tr \log [D(\bar U)+ R^\dagger \dot R]
= N_c \Tr \log [D(\bar U)]
+ \Theta^{sea} \int d\tau \Sp\, (R^\dagger \dot R)^2 + \ldots\,,
\label{moment-of-inertia-definition}
\end{equation}
and
\begin{equation}
\frac 1{D(\bar U)+R^\dagger \dot R}=\frac 1{D(\bar U)}
-\frac 1{D(\bar U)} R^\dagger\dot R  \frac1{D(\bar U)}
+\ldots\,.
\label{Dexpansion}
\end{equation}
In eq.(\ref{moment-of-inertia-definition}) the second term of the action is
a
function of the angular velocity only. As we will see later, the coefficient
$\Theta^{sea}$ is the Dirac sea contribution to the  moment of inertia of
the soliton.

Using the expansions (\ref{moment-of-inertia-definition}),(\ref{Dexpansion}),
in eqs.~(\ref{sea-contributionR}), (\ref{valence-contributionR}) up to
terms quadratic in $R^\dagger\dot R$ we arrive at
a functional integral over the time dependent orientation matrices $R(x_4)$
with the action quadratic in angular velocities~\cite{Diakonov89}.
In large $N_c$ limit, the latter corresponds to the hamiltonian of the
quantum spherical rotator:
\begin{equation}
H_{rot} = J_a^2/(2\Theta)\,,
\label{h-rot}
\end{equation}
It means that despite of the fact that in general the path integral over $R$
runs over all possible trajectories, in large $N_c$ limit the main
contribution comes from trajectories close to those of the quantum rotator
with hamiltonian~(\ref{h-rot}). In eq.~(\ref{h-rot}) $J_a$ is the spin
operator of the nucleon and
\begin{equation}
\Theta=\Theta^{sea}+\Theta^{val}
\label{Theta}
\end{equation}
is the total moment of inertia, including also the valence quark
contribution $\Theta^{val}$ --- see eq. (\ref{moment-of-inertia-result})
below. According to eq.~(\ref{h-rot}) the quantization rule (for the
Euclidean time) is given by
\begin{equation} \Sp(R^\dagger \dot R \tau_a) \to \frac{J_a}{\Theta} \,,
\label{quantization-rule}
\end{equation}
We also make use of the identity
\begin{equation}
R^\dagger \dot R = \frac 12\Sp(R^\dagger \dot R \tau_a)\,\tau^a
\end{equation}
in order to separate the part, acting on the rotational functions of the
nucleon $\psi_N(R)$, from the Pauli matrices
$\tau^a$ which act in the isospin space of the quark states.

It should be also noted that in expansions
(\ref{moment-of-inertia-definition}) and (\ref{Dexpansion})
we have neglected the terms containing  derivatives of
angular velocity $\partial_\tau (R^\dagger\dot R)$. Such a treatment is
consistent in a sense that for the hamiltonian~(\ref{h-rot})
we have
\begin{equation}
\partial_\tau(R^\dagger\dot R)=[H_{rot},R^\dagger\dot R]= 0
\end{equation}
for any rotation state of $H_{rot}$.

After quantization (\ref{quantization-rule})
we have to deal with not-commuting collective operators.
Apparently, since the path integral corresponds to the following operator
construction:
\begin{equation}
\int\limits_{R(T_1)=R_1}^{R(T_2)=R_2}{\cal D}R\ F_1(R(t_1))\cdots
F_n(R(t_n))\,\e^{-S(R)}=\bra{R_2,T_2}\,{\cal T}\{\hat
F_1(R(t_1))\ldots\hat F_n(R(t_n))\}\,\ket{R_1,T_1} \,,
\label{equivalence:integral-operators}\end{equation}
where ${\cal T}$ stands for time-ordering of operators $\hat F_n(R(t_n))$
in the Heisenberg representation, the order of the collective operators,
which appear after the integration over $R$, is strictly fixed
by the time ordering.

Back to the sea contribution (\ref{sea-contributionR}),
expanding both the action and the integrand
in powers of $R^\dagger\dot R$ and performing the functional integration
over time dependent $R(x_4)$ in large $N_c$ limit,
we now have
\begin{equation}
\langle N^\prime, {\vec p}^\prime | \Psi^\dagger O \Psi | N, {\vec p}
\rangle^{sea} = N_c\int \d^3x\, \e^{i({\vec p}^\prime - {\vec p}) {\vec x}}
 \int \d R \, \psi_{N^\prime}^\ast (R)
\Sp \Bigl[ R^\dagger O R \langle \, {\vec x},0 |
\frac{-1}{D(\bar U) + R^\dagger \dot R } | {\vec x},0 \rangle
\Bigr]\Bigg\vert_{\Sp(R^\dagger \dot R\tau_a){\to\atop{\cal T}} J_a
/\Theta} \psi_N(R) \,, \label{sea-contribution-2}
\end{equation}
where ${\to\atop{\cal T}}$ means again time-ordering of the collective
operators. The rotational wave functions $\psi_N(R)$
comes from the product of $N_c$ matrices $R(T/2)$, which appear in eq.
(\ref{sea-contribution})  after transformation (\ref{Eq10b}),
contracted by the matrices $\Gamma_N$ from the definition of the baryon
current (\ref{baryon-current}). For given spin
$J,J_3$ and isospin $T,T_3$ these wave functions can be expressed
through the Wigner $D$ functions
\begin{equation}
\psi_{TT_3JJ_3}(R) = (-1)^{T+T_3}\sqrt{2T+1}\,D^{T=J}_{-T_3,J_3}(R)\,.
\label{D-functions}
\end{equation}
In eq.(\ref{sea-contribution-2}) the integral over ${\vec x}$ has appeared
due to the integration over the translational modes of the soliton and
the matrix element in the integrand is
given by
\begin{eqnarray}
&&\Sp\Bigl[ R^\dagger O R(0) \, \langle {\vec x},0 |
\frac{-1}{D(\bar U) + R^\dagger \dot R } | {\vec x},0 \rangle
\Bigr]\Bigg\vert_{\Sp(R^\dagger \dot R\tau_a){\to\atop{\cal
T}}J_a/\Theta}\,=\,
- \Sp \left[ R^\dagger O R \, \langle  {\vec x},0 |
\frac{1}{D(\bar U)} | {\vec x},0 \rangle \right]\nonumber\\
&&+ \frac{1}{2\Theta} \Sp \int d^4y \bigl[
\theta(y_4) J_a \, R^\dagger O R
+ \theta(-y_4) \, R^\dagger O R J_a\bigr] \,
 \langle {\vec x},0 | \frac{1}{D(\bar U)}  | y \rangle  \tau^a
\langle y |\frac{1}{D(\bar U)} | {\vec x},0 \rangle \,.
\label{sea-contribution-3}
\end{eqnarray}

As a next step we use the spectral representation for the quark propagator
\begin{equation}
\langle x^\prime|\frac{1}{D(\bar U)}|  x \rangle
= \theta(x_4^\prime-x_4) \sum\limits_{\varepsilon_n>0}
\e^{-\varepsilon_n(x_4^\prime-x_4)}\,\Phi_n({\vec x}^\prime)  \,
\Phi_n^\dagger(\vec x)
- \theta(x_4-x_4^\prime) \sum\limits_{\varepsilon_n<0}
\e^{-\varepsilon_n(x_4^\prime-x_4)}\,\Phi_n({\vec x}^\prime)  \,
\Phi_n^\dagger(\vec x)
\label{spectral-representation}
\end{equation}
written in terms of eigenvalues $\varepsilon_n$ and eigenfunctions $\Phi_n$
of the Dirac hamiltonian (\ref{HAMIL})
\begin{equation}
h(\bar U) \Phi_n = \varepsilon_n \Phi_n\,.
\end{equation}
in order to evaluate the matrix elements and the integral over $y_4$
in (\ref{sea-contribution-3}). Finally we get for the sea contribution
(\ref{sea-contribution}) the following result:
\begin{eqnarray}
&& \langle N^\prime, {\vec p}^\prime | \Psi^\dagger O \Psi | N, {\vec
p} \rangle^{sea}
= N_c
\int \d^3 x \e^{i({\vec p}^\prime - {\vec p}) {\vec x}} \int \d R \,
\psi_{N^\prime}^\ast(R)
\Biggl\{ \sum\limits_{\varepsilon_n<0}
\Phi_n^\dagger({\vec x}) \, R^\dagger O R \,\Phi_n({\vec x})
+\frac{1}{2\Theta}\sum\limits_{m,n}\frac{1}
{\varepsilon_n-\varepsilon_m}\nonumber\\
&&\times\Bigl\{ - \theta(\varepsilon_n) \, \theta(-\varepsilon_m)
J_a\left[\Phi_n^\dagger({\vec x}) \, R^\dagger O R \,\Phi_m({\vec
x})\right] + \theta(-\varepsilon_n) \, \theta(\varepsilon_m) \,
\left[\Phi_n^\dagger({\vec x}) \, R^\dagger O R \,\Phi_m({\vec x})\right] \,
 J_a \Bigr\}\,
\times\langle m |\tau^a| n \rangle  \Biggr\}
\psi_N(R)
\label{sea-contribution-result}
\end{eqnarray}

Now the valence level contribution to the form factor
(\ref{valence-contribution}) is in order. Note that in the limit of large
Euclidean time $T \to + \infty$ the quark propagator
(\ref{spectral-representation}) is dominated by the contribution
of the lowest level with $\varepsilon_n>0$, i.e. by the valence quark level
with the wave function $\Phi_{val}({\vec x})$. We perform analogous
calculations as in the case of Dirac sea, i.e. using the  rotating ansatz
(\ref{rotating-soliton}) in (\ref{valence-contribution}) and expanding up to
the linear order in the angular velocity with time-ordering of collective
operators we arrive at the following result for the valence  quark
contribution:
\begin{eqnarray}
&&  \langle N^\prime, {\vec p}^\prime | \Psi^\dagger O \Psi | N, {\vec
p} \rangle^{val}
= N_c\int \d^3 x\e^{i({\vec p}^\prime - {\vec p}) {\vec x}}
\int\d R \, \psi_{N^\prime}^\ast(R)
\Biggl\{
\Phi_{val}^\dagger({\vec x}) \, R^\dagger O R \,\Phi_{val}({\vec
x})\nonumber\\
&&+ \frac{1}{2\Theta} \sum\limits_{\varepsilon_n\neq \varepsilon_{val}}
 \frac{1}{\varepsilon_{val} - \varepsilon_n}
\Bigl\{ \theta(\varepsilon_n)\Bigl[  \, J_a \,
[\Phi_n^\dagger({\vec x}) \, R^\dagger O R \,\Phi_{val}({\vec x})]
\langle val |\tau^a| n \rangle
+ [\Phi_{val}^\dagger({\vec x}) \, R^\dagger O R \,\Phi_n({\vec x})] \,
\langle n |\tau^a| val \rangle J_a\Bigr]
\nonumber \\
&&
+  \theta(-\varepsilon_n) \Bigl[  J_a \,
[\Phi_{val}^\dagger({\vec x}) \, R^\dagger O R \,\Phi_n({\vec x})] \,
\langle n |\tau^a| val \rangle
+ [\Phi_n^\dagger({\vec x}) \, R^\dagger O R \,\Phi_{val}({\vec x})] \,
\langle val |\tau^a| n \rangle \, J_a \Bigr]\Bigr\} \Biggr\}
\psi_N(R)
\label{valence-contribution-result}
\end{eqnarray}

Finally we add the valence contribution (\ref{valence-contribution-result})
to the sea part (\ref{sea-contribution-result}):
\begin{eqnarray}
&&  \langle N^\prime, {\vec p}^\prime | \Psi^\dagger O \Psi | N, {\vec p}
\rangle
= N_c
\int \d^3 x
\e^{i({\vec p}^\prime - {\vec p}) {\vec x}}
 \int \d R \, \psi_{N^\prime}^\ast(R)
\Biggl\{ \sum\limits_{\varepsilon_n \leq \varepsilon_{val}}
\Phi_n^\dagger({\vec x}) \, R^\dagger O R \,\Phi_n({\vec x})
\nonumber\\
&&- \frac{1}{2\Theta} \sum\limits_{\varepsilon_n>\varepsilon_{val} \atop
\varepsilon_m\leq \varepsilon_{val}}
 \frac{1}{\varepsilon_n - \varepsilon_m}
\Bigl\{ J_a [\Phi_n^\dagger({\vec x}) \, R^\dagger O R \,\Phi_m({\vec
x})] \, \langle m |\tau^a| n \rangle
 + [\Phi_m^\dagger({\vec x}) \, R^\dagger O R \,\Phi_n({\vec x})] \,
\langle n |\tau^a| m \rangle \, J_a \Bigr\} \Biggr\}
\psi_N(R)
\label{form-factor-result}
\end{eqnarray}
It is easy to see that the result has the same structure as
the sea contribution but with the valence quark included
into the occupied states.

\section{Regularization}

The NJL model is not renormalizable. Even in
the leading order of the saddle point approximation
one faces ultraviolet divergences. The integrand in the r.h.s.
of eq.~(\ref{sea-contribution}) contains fermionic determinant
Det$[\,D(U)]$ and a matrix element:
\begin{equation}
\Sp O \langle 0 |  \frac{-1}{D(U)} | 0 \rangle
=\frac{\delta}{\delta\epsilon(0)}\Tr\log[D(U)-\epsilon O]\Bigr|_{\epsilon=0}
\label{epsilon-source}
\end{equation}
both divergent. Apparently, in order to make it finite it is sufficient to
regularize
\begin{equation}
\Tr\log D_\epsilon\equiv\Tr\log[D(U)-\epsilon O] = \Tr\log[D(\bar U) +
R^\dagger\dot R - \epsilon R^\dagger O R] \,. \end{equation}

In fact, we are interested in a particular operator $O$, for which only the
real part of $\Tr\log D_\epsilon$ is divergent. The corresponding
regularized sea contribution to the matrix element (\ref{sea-contribution})
has the form
\begin{equation}
\langle N^\prime, {\vec p}^\prime | \Psi^\dagger O \Psi | N, {\vec p}
\rangle_{reg}^{sea}
= N_c
\int \d^3 x
\e^{i({\vec p}^\prime - {\vec p}) {\vec x}}
 \int\d R \, \psi_{N^\prime}^\ast(R)
 \Bigl[ \frac{\delta}{\delta\epsilon({\vec x},0)}
\mbox{Re} \, (\Tr\log D_\epsilon)_{reg}
\Bigr]\Bigg\vert_{\Sp(R^\dagger \dot R\tau_a){\to\atop{\cal T}}J_a/\Theta}
\psi_N(R) \,, \label{sea-contribution-regularized}
\end{equation}
where we used a proper-time regularization
\begin{eqnarray}
\mbox{Re}(\Tr \log D_\epsilon)_{reg}&\equiv&
\frac 12 [\Tr\log(D_\epsilon^\dagger D_\epsilon)]_{reg}
\nonumber\\
&=&-\frac 12 \int\limits_{1/\Lambda^2}^{\infty}\frac{du}{u} \, \Tr
\left[ \e^{-uD_\epsilon^\dagger D_\epsilon}
-\e^{-uD_0^\dagger D_0}
\right] \,.
\label{proper-time-regularization}
\end{eqnarray}
Hence
\begin{eqnarray}
\frac{\delta}{\delta\epsilon(x)} \mbox{Re} \,
(\Tr\log D_\epsilon)_{reg}
&=& -\frac 12\, \Sp \, R^\dagger(x_4)OR(x_4)
\langle x | \e^{-D^\dagger D/\Lambda^2} D^{-1} | x \rangle
\nonumber\\
&&
 - \frac 12 \, \Sp \, R^\dagger(x_4)O^\dagger R(x_4)
\langle x |(D^\dagger)^{-1} \e^{-D^\dagger D/\Lambda^2} | x \rangle \, .
\label{variational-derivative-regularized}
\end{eqnarray}
Here D stands for
\begin{equation}
D=D_{\epsilon=0} = \partial_\tau + h(\bar U)
+ R^\dagger\dot R \,
\end{equation}
Note that we assume $\epsilon(x)$ to be real but
the matrix $O$ is not necessarily hermitean.

Using the expansion
\begin{equation}
D^{-1}=\frac{1}{\partial_\tau+h}
-\frac{1}{\partial_\tau+h} R^\dagger\dot R  \frac{1}{\partial_\tau+h}
+\ldots
\end{equation}
up to the linear order in the angular velocity we get
\begin{eqnarray}
&&\e^{-D^\dagger D/\Lambda^2} = \e^{(\partial_\tau^2-h^2)/\Lambda^2}
+\frac{1}{\Lambda^2} \int\limits_0^1\d\alpha \,
\e^{\alpha(\partial_\tau^2-h^2)/\Lambda^2}(\partial_\tau-h) \,
R^\dagger\dot R \,  \e^{(1-\alpha)(\partial_\tau^2-h^2)/\Lambda^2}
\nonumber\\
&&
+\frac{1}{\Lambda^2} \int\limits_0^1\d\alpha \,
\e^{\alpha(\partial_\tau^2-h^2)/\Lambda^2} \,
R^\dagger\dot R \,  \e^{(1-\alpha)(\partial_\tau^2-h^2)/\Lambda^2}
(\partial_\tau+h) + \ldots
\end{eqnarray}
After quantization
(\ref{quantization-rule}) of the time-ordered collective
operators and using the spectral representation for the
quark propagator (\ref{spectral-representation}) we get
for the regularized Dirac sea contribution
(\ref{sea-contribution-regularized})in the form
\begin{eqnarray}
&&  \langle N^\prime, {\vec p}^\prime | \Psi^\dagger O \Psi | N, {\vec p}
\rangle_{reg}^{sea}
= N_c
\int \d^3 x
\e^{i({\vec p}^\prime - {\vec p}) {\vec x}}
 \int dR \, \psi_{N^\prime}^\ast(R)
\biggl\{ \sum\limits_n {\cal R}_1(\varepsilon_n,\eta) \,
\Phi_n^\dagger({\vec x}) \, R^\dagger O R \,\Phi_n({\vec x})
\nonumber\\
&&
+ \frac{1}{4\Theta} \sum\limits_{m,n}
\Bigl[ {\cal R}_2^{(+)}(\varepsilon_m,\varepsilon_n,\eta)  J_a \,
[\Phi_m^\dagger({\vec x}) \, R^\dagger O R \,\Phi_n({\vec x})] \,
 +  {\cal R}_2^{(-)}(\varepsilon_m,\varepsilon_n,\eta)
[\Phi_m^\dagger({\vec x}) \, R^\dagger O R \,\Phi_n({\vec x})] \,
 J_a \Bigr] \,\langle n |\tau^a| m \rangle      \biggr\}
\psi_N(R) \, .
\label{sea-regularized}
\end{eqnarray}
The coefficient $\eta=1$ for hermitean matrix $O$ and $\eta=-1$ for
anti-hermitean one:
\begin{equation}
O^\dagger = \eta O \,.
\label{eta-definition}
\end{equation}
The regularization functions are given by
\begin{equation}
{\cal R}_1(\varepsilon_n,\eta)=-\,\frac{1+\eta}{2} \,\varepsilon_n
\int\limits_{-\infty}^{\infty}\frac{\d\omega}{2\pi} \>
\frac{\e[-(\omega^2+\varepsilon_n^2)/\Lambda^2]}{\omega^2+\varepsilon_n^2}
\, ,
\end{equation}
\begin{equation}
{\cal R}_2^{(\pm)}(\varepsilon_m,\varepsilon_n,\eta)
={\cal R}^{(\pm)}(\varepsilon_m,\varepsilon_n)
- \eta {\cal R}^{(\pm)}(\varepsilon_n,\varepsilon_m) \, ,
\end{equation}
\begin{eqnarray}
&&{\cal R}^{(\pm)}(\varepsilon_m,\varepsilon_n) =
\int\limits_{-\infty}^{\infty}\frac{\d\omega}{2\pi}
\int\limits_{-\infty}^{\infty}\frac{\d\omega^\prime}{2\pi} \>
 \frac{1}{\pm i(\omega - \omega^\prime) +0}
\frac{1}{(i\omega + \varepsilon_n)(i\omega^\prime + \varepsilon_m)}
\nonumber\\
&&-\frac{1}{\Lambda^2}
\left[1+\frac{i\omega - \varepsilon_n}{i\omega^\prime + \varepsilon_m}  \right]
\int_0^1\d\alpha
\e\left[-\frac{\alpha(\omega^2+\varepsilon_n^2)
+(1-\alpha)({\omega^\prime}^2+\varepsilon_m^2) }{\Lambda^2} \right]
\biggr\} \, .
\end{eqnarray}

\section{Electromagnetic form factors}

In this section we apply the formalism, developed
above, to the case of the electromagnetic form factors.
We pay a special attention to the ultraviolet regularization
to preserve the electromagnetic gauge invariance. Following this
regularization prescription  we evaluate the nucleon matrix
elements of the electromagnetic current in the NJL model
and relate them to the Sachs form factors.

In the previous section we have presented a scheme for evaluating the
regularized determinant with a small external source $\epsilon O$
(\ref{epsilon-source}). In the case of the electromagnetic form factors
this external source is an electromagnetic field $A_\mu$:
\begin{equation}
D(U) - \epsilon O \quad \to \quad
D(U,QA_\mu) = \gamma_4[\gamma_\mu(\partial_\mu - i Q A_\mu) + MU^{\gamma_5}]
\, .
\end{equation}
Here $Q$ is the quark charge matrix
\begin{equation}
Q=\left( \begin{array}{cc}
         2/3  & 0 \\
         0  & -1/3     \end{array}\right) = \frac{1}{6} + \frac{\tau^3}{2}
\, .
\label{quark-charge-matrix}
\end{equation}

Our proper-time regularization in the Euclidean space
(\ref{proper-time-regularization}) includes operator $D^\dagger D$.
Therefore, in order to preserve the gauge invariance
in (\ref{proper-time-regularization})
the Euclidean electromagnetic field $A_\mu$ should be real.

The complex conjugate Euclidean
Dirac matrices $\gamma_\mu^\ast$ are  connected
to $\gamma_\mu$ by some unitary transformation $V$
\begin{equation}
\gamma_\mu^\ast = V \gamma_\mu V^{-1}  \qquad(\mu=1,2,3,4,5)\,.
\end{equation}
It is easy to show that
\begin{equation}
(V\tau^2) D(U,QA_\mu) (V\tau^2)^{-1} = [D(U,Q^\prime A_\mu)]^\ast\,,
\label{D-D-ast}
\end{equation}
where
\begin{equation}
Q^\prime =  - \frac{1}{6} + \frac{\tau^3}{2}
\label{Q-prime}
\end{equation}
and the asterisk stands for the complex (not hermitean!) conjugation.

Comparing $Q$ (\ref{quark-charge-matrix}) and $Q^\prime$ (\ref{Q-prime})
we see that their isovector part remains the same whereas the isoscalar
component changes the sign. The identity  (\ref{D-D-ast}) shows that the
isoscalar electromagnetic form factors originate from the
imaginary part of $\Tr\log D(U,QA_\mu)$ whereas the isovector form factors
are generated by the real part of $\Tr\log D(U,QA_\mu)$.

Since the imaginary part of $\Tr\log D(U,QA_\mu)$ is ultraviolet finite,
for the isoscalar form factors we use directly
the non-regularized expression (\ref{form-factor-result}).
In this case the
integral over $R$ becomes trivial and reduces to the orthogonality
condition for the rotational wave functions or to the standard
matrix elements of the spin operator $J_a$
\begin{eqnarray}
&&  \langle T_3^\prime,J_3^\prime, {\vec p}^\prime | \Psi^\dagger i
\gamma_4\gamma_\mu  \Psi
| T_3,J_3, {\vec p} \rangle
= N_c\, \delta_{T_3^\prime T_3}
\int \d^3 x
\e^{i({\vec p}^\prime - {\vec p}) {\vec x}}
\Biggl\{\delta_{J_3^\prime J_3} \sum\limits_{\varepsilon_n \leq
\varepsilon_{val}}
\Phi_n^\dagger({\vec x}) \, i \gamma_4\gamma_\mu \,\Phi_n({\vec
x})\nonumber\\ &&- \frac{1}{4\Theta}\, {(\tau^a)}_{J_3^\prime J_3}
\sum\limits_{\varepsilon_n>\varepsilon_{val} \atop \varepsilon_m\leq
\varepsilon_{val}}  \frac{1}{\varepsilon_n - \varepsilon_m}
\Bigl[ [\Phi_n^\dagger({\vec x}) \, i \gamma_4\gamma_\mu
\,\Phi_m({\vec x})] \, \langle m |\tau^a| n \rangle
 + [\Phi_m^\dagger({\vec x}) \, i \gamma_4\gamma_\mu  \,\Phi_n({\vec x})] \,
\langle n |\tau^a| m \rangle  \Bigr] \Biggr\}\,.
\label{isoscalar-form-factor}
\end{eqnarray}

In the case of the isovector current, however, we have to use
the regularized expression (\ref{sea-regularized}) for the sea contribution,
which leads to: \begin{eqnarray}
&&  \langle N^\prime, {\vec p}^\prime | \Psi^\dagger  i \gamma_4\gamma_\mu
\tau^3
 \Psi | N, {\vec p}
\rangle_{reg}^{sea}
= N_c
\int \d^3 x
\e^{i({\vec p}^\prime - {\vec p}) {\vec x}}
 \int dR \, \psi_{N^\prime}^\ast(R)
\biggl\{D_{3b}(R)\sum\limits_n {\cal R}_1(\varepsilon_n,\eta) \,
\Phi_n^\dagger({\vec x}) \,  i \gamma_4\gamma_\mu   \tau^b \,
\Phi_n({\vec x})
\nonumber\\
&&
+ \frac{1}{4\Theta} \sum\limits_{m,n}
\Bigl[ {\cal R}_2^{(+)}(\varepsilon_m,\varepsilon_n,\eta)  J_a \, D_{3b}(R)
 +  {\cal R}_2^{(-)}(\varepsilon_m,\varepsilon_n,\eta)   D_{3b}(R)\, J_a
\Bigr]
[\Phi_m^\dagger({\vec x}) \,  i \gamma_4\gamma_\mu   \tau^b \,
\Phi_n({\vec x})] \,
  \langle n |\tau^a| m \rangle      \biggr\}
\psi_N(R)\,,
\label{isovector-form-factor}
\end{eqnarray}
where
\begin{equation}
D_{ab}(R) = \frac 12 \, \Sp ( R^\dagger \tau^a R \tau^b)\,.
\end{equation}

Note that $i \gamma_4\gamma_\mu \tau^b$ is hermitean for $\mu=1,2,3$
and anti-hermitean for $\mu=4$ which means that the coefficient $\eta$
(defined in (\ref{eta-definition}))  has the following values:
\begin{equation}
\eta
=\left\{ \begin{array}{ll}
         1   &\mbox{if }   \mu=1,2,3  \\
         -1  & \mbox{if }   \mu=4 .   \end{array} \right.
\label{eta-definition-2}
\end{equation}

Now the rotational matrix elements in (\ref{isovector-form-factor})
can be easily computed using
\begin{equation}
\langle T_3^\prime J_3^\prime  |J_{a}| T_3 J_3 \rangle = \frac 12
\,\delta_{T_3^\prime T_3}\,(\tau^a)_{J_3^\prime J_3} \, ,
\end{equation}
\begin{equation}
\langle T_3^\prime J_3^\prime  |D_{ab}(R)| T_3 J_3 \rangle = - \frac 13 \,
(\tau^a)_{T_3^\prime T_3}\, (\tau^b)_{J_3^\prime J_3} \, ,
\end{equation}
 From eq.(\ref{isovector-form-factor}) we find:
\begin{eqnarray}
&&  \langle T_3^\prime,J_3^\prime, {\vec p}^\prime | \Psi^\dagger  i
\gamma_4\gamma_\mu \tau^3
 \Psi | T_3,J_3, {\vec p} \rangle_{reg}^{sea}
= N_c T_3 \delta_{T_3^\prime T_3}\,
\int \d^3 x
\e^{i({\vec p}^\prime - {\vec p}) {\vec x}}
\biggl\{- \, \frac{2}{3} \,(\tau^b)_{J_3^\prime J_3}
\sum\limits_n {\cal R}_1(\varepsilon_n,\eta) \,
\Phi_n^\dagger({\vec x}) \,  i \gamma_4\gamma_\mu   \tau^b \,
\Phi_n({\vec x})
\nonumber\\
&&
- \frac{1}{12\Theta}  \sum\limits_{m,n}
\Bigl\{ \delta_{ab} \delta_{J_3^\prime J_3}
\Bigl[ {\cal R}_2^{(+)}(\varepsilon_m,\varepsilon_n,\eta)
 + {\cal R}_2^{(-)}(\varepsilon_m,\varepsilon_n,\eta)
 \Bigr]
+i \epsilon_{abc} (\tau^c)_{J_3^\prime J_3}
\Bigl[ {\cal R}_2^{(+)}(\varepsilon_m,\varepsilon_n,\eta)
 -  {\cal R}_2^{(-)}(\varepsilon_m,\varepsilon_n,\eta)
 \Bigr]\Bigr\}
\nonumber\\
&&
\times[\Phi_m^\dagger({\vec x}) \,  i \gamma_4\gamma_\mu   \tau^b  \,
\Phi_n({\vec x})] \,
\langle n |\tau^a| m \rangle      \biggr\} \,.
\label{isovector-form-factor-3}
\end{eqnarray}

Under the transformation $V\tau^2$ (\ref{D-D-ast}) we have
\begin{equation}
(V\tau^2) \gamma_\mu (V\tau^2)^{-1} = \gamma_\mu^\ast = \gamma_\mu^T,
\quad
(V\tau^2) \tau^a (V\tau^2)^{-1} = - (\tau^a)^\ast = -(\tau^a)^T,
\end{equation}
\begin{equation}
(V\tau^2) h (V\tau^2)^{-1} = h^\ast = h^T\,,
\end{equation}
where the superscript $T$ stands for the transposition in both the
matrix indices and the coordinate space.
The last equation shows that the eigenfunctions of $h$ can be chosen in
such a way that
\begin{equation}
(V\tau^2) \Phi_m({\vec x}) = \Phi_m^\ast({\vec x})
\end{equation}
and hence, the following relations hold:
\begin{equation}
\Phi_n^\dagger({\vec x}) \gamma_4\gamma_\mu \Phi_m({\vec x})langle m |
\tau^a| n \rangle
= \eta\, \Phi_m^\dagger({\vec x})  \gamma_4\gamma_\mu \Phi_n({\vec
x})\langle n | \tau^a| m \rangle  \,,
\label{gamma-transposition}
\end{equation}
\begin{equation}
\Phi_n^\dagger({\vec x}) \gamma_4\gamma_\mu\tau^a \Phi_m({\vec
x})\langle m | \tau^a| n \rangle
 = -\eta\, \Phi_m^\dagger({\vec x}) \gamma_4\gamma_\mu\tau^a \Phi_n({\vec
x})\langle n | \tau^a| m \rangle \label{gamma-tau-transposition}
\end{equation}
with values for $\eta$ are given in (\ref{eta-definition-2}).

Using these symmetry properties of matrix elements
we can simplify expression~(\ref{isoscalar-form-factor})
\begin{eqnarray}
&&  \langle T_3^\prime,J_3^\prime, {\vec p}^\prime | \Psi^\dagger i
\gamma_4\gamma_\mu  \Psi
| T_3,J_3, {\vec p} \rangle
= N_c \delta_{T_3^\prime T_3}
\int \d^3 x
\e^{i({\vec p}^\prime - {\vec p}) {\vec x}}
\Biggl\{\delta_{J_3^\prime J_3}
\, \frac{1-\eta}{2} \,
 \sum\limits_{\varepsilon_n \leq \varepsilon_{val}}
\Phi_n^\dagger({\vec x}) \, i \gamma_4\gamma_\mu \,\Phi_n({\vec x})
\nonumber\\
&&
-\, \frac{1+\eta}{4\Theta}\, {(\tau^a)}_{J_3^\prime J_3}
\sum\limits_{\varepsilon_n>\varepsilon_{val} \atop \varepsilon_m\leq
\varepsilon_{val}}
 \frac{1}{\varepsilon_n - \varepsilon_m}
 \,
\Phi_n^\dagger({\vec x}) \, i \gamma_4\gamma_\mu \,\Phi_m({\vec x}) \,
\langle m |\tau^a| n \rangle
   \Biggr\} \,.
\label{isoscalar-form-factor-2}
\end{eqnarray}

Starting from the general formula (\ref{valence-contribution-result})
analogous calculations of the non-regularized valence
quark contribution to the matrix element of the isovector part of the
electromagnetic current can be performed:
\begin{eqnarray}
&&  \langle T_3^\prime,J_3^\prime  , {\vec p}^\prime |
 \Psi^\dagger i \gamma_4\gamma_\mu \tau^3 \Psi
| T_3, J_3  , {\vec p} \rangle^{val}
= N_c\, T_3 \,\delta_{T_3^\prime T_3}
\int \d^3 x
\e^{i({\vec p}^\prime - {\vec p}) {\vec x}}
\Biggl\{ -\frac{1+\eta}{3}  (\tau^b)_{J_3^\prime J_3}
\Phi_{val}^\dagger({\vec x}) \,  i \gamma_4\gamma_\mu \tau^b
 \,\Phi_{val}({\vec x})
\nonumber\\
&&
- \,\frac{1}{6\Theta} \,
   \sum\limits_{\varepsilon_n\neq \varepsilon_{val}}
\frac{1}{\varepsilon_{val} - \varepsilon_n}
\bigl[ \theta(\varepsilon_n) - \eta \theta(-\varepsilon_n)\bigr]
  [(1-\eta) \delta_{ab} \delta_{J_3^\prime J_3}
  + (1+\eta)i \epsilon_{abc} (\tau^c)_{J_3^\prime J_3}]
\nonumber\\
&&
\times[\Phi_n^\dagger({\vec x}) \,  i \gamma_4\gamma_\mu \tau^b
 \,\Phi_{val}({\vec x})] \,
\langle val |\tau^a| n \rangle \Biggr\}
\label{valence-contribution-result-isovector} \,.
\end{eqnarray}
Up to now we worked with Euclidean Dirac matrices. Since the same
combination $i\gamma_4\gamma_\mu$ enters both left and right
hand sides of our formulas for form factors the transition
to the Minkowskian Dirac matrices is trivial --- it is sufficient to
replace the Euclidean $i\gamma_4\gamma_\mu$ by Minkowskian
$i\gamma^0\gamma^\mu$  everywhere in eqs.
(\ref{isovector-form-factor-3}),
(\ref{isoscalar-form-factor-2}) and
(\ref{valence-contribution-result-isovector}).
Henceforward, we work in the Minkowski space-time.

The nucleon electromagnetic Sachs form factors are related to the matrix
element of the electromagnetic current
\begin{equation}
j^\mu = \bar\Psi\gamma^\mu Q\Psi = \Psi^\dagger\gamma^0\gamma^\mu
\left(\frac16 + \frac12\tau^3\right)\Psi
\end{equation}
in the standard way:
\begin{equation}
\langle J_3^\prime, p^\prime | j^0(0) | J_3,p \rangle = G_E(q^2)
\delta_{J_3^\prime J_3} \,, \label{Sachs-electric-form-factor}
\end{equation}
\begin{equation}
\langle J_3^\prime, p^\prime | j^k(0) | J_3,p \rangle = \frac{i}{2{\cal M}_N}
\epsilon^{klm} (\tau^l)_{J_3^\prime J_3} q^m
 G_M(q^2)\,,
\label{Sachs-magnetic-form-factor}
\end{equation}
where
\begin{equation}
q=p^\prime-p \,.
\end{equation}
The isoscalar and isovector parts of the form factors
are defined by
\begin{equation}
G_{E(M)} = \frac12 G_{E(M)}^{T=0} + T_3 G_{E(M)}^{T=1} \,.
\label{isoscalar-isovector}
\end{equation}
Using our result (\ref{isoscalar-form-factor-2}) for the nucleon matrix
elements of the isoscalar vector current and definition
(\ref{Sachs-electric-form-factor}) we find for the isoscalar electric form
factor a simple result:
\begin{eqnarray}
  G_E^{T=0}(q^2)= \frac{N_c}{3}\int \d^3 x
\e^{i{\vec q} {\vec x}} \Bigl\{\sum\limits_{\varepsilon_n \leq
\varepsilon_{val}} \Phi_n^\dagger({\vec x}) \Phi_n({\vec x})
- \sum\limits_{\varepsilon_n^{(0)} <0}
\Phi_n^{(0)\dagger}({\vec x}) \Phi_n^{(0)}({\vec x}) \Bigr\}
\label{form-factor-electric-isosinglet}
\end{eqnarray}
Here we have to subtract the vacuum contribution which  provides the
correct normalization:
\begin{eqnarray}
  G_E^{T=0}(0)=1
\end{eqnarray}

For the electric isovector form factor
from(\ref{isovector-form-factor-3}) and
(\ref{valence-contribution-result-isovector}) we obtain:
\begin{eqnarray}
&&
  G_E^{T=1}(q^2)
=  \frac{N_c}{6\Theta} \,
\int \d^3 x
\e^{i({\vec p}^\prime - {\vec p}) {\vec x}}
 \biggl\{
   \sum\limits_{m,n}
 {\cal R}_\Theta^\Lambda(\varepsilon_m,\varepsilon_n)
[\Phi_m^\dagger({\vec x}) \,   \tau^a  \, \Phi_n({\vec x})] \,
\langle n |\tau^a| m \rangle
\nonumber\\
&&
-  \sum\limits_{\varepsilon_n\neq \varepsilon_{val}}
\frac{1}{\varepsilon_{val} - \varepsilon_n}
[\Phi_n^\dagger({\vec x}) \,   \tau^a  \,
\Phi_{val}({\vec x})] \,
\langle val |\tau^a| n \rangle
\biggl\}\,,
\label{electric-isovector-form-factor}
\end{eqnarray}
where the regulator has the form:
\begin{eqnarray}
{\cal R}_\Theta^\Lambda(\varepsilon_m,\varepsilon_n)
&=& - \frac{1}{4}
 \left[{\cal R}_2^{(+)}(\varepsilon_m,\varepsilon_n,-1)
 +  {\cal R}_2^{(-)}(\varepsilon_m,\varepsilon_n,-1)\right]
\nonumber\\
&=&\frac{1}{4\sqrt{\pi}} \int\limits_{1/\Lambda^2}^{\infty}
\frac{du}{\sqrt{u}}
\left(
\frac{1}{u} \frac
{ \e^{-u\varepsilon_n^2} - \e^{-u\varepsilon_m^2} }
{\varepsilon_m^2 - \varepsilon_n^2 }
- \frac{\varepsilon_n \e^{-u\varepsilon_n^2}
+\varepsilon_m \e^{-u\varepsilon_m^2} }
{\varepsilon_m + \varepsilon_n }
\right) \,.
\end{eqnarray}

It can be easily seen that the calculation of the moment
of inertia is very similar to the case of the electric isovector form
factor. Indeed, the Dirac sea contribution to the moment of
inertia (\ref{moment-of-inertia-definition})
can be written in the form:
\begin{equation}
\Theta^{sea} \delta_{ab} \int\d\tau =
\frac{N_c }{4}
\frac{\partial^2}{\partial\epsilon_a\partial\epsilon_b}
\mbox{Re} \Tr\log (D_\epsilon^\dagger D_\epsilon)_{reg}\,,
\label{moment-of-inertia-2}
\end{equation}
where now
\begin{equation}
D_\epsilon  =  D(\bar U) - i \epsilon_c\tau^c     \,.
\end{equation}
Note that the first $\epsilon$ derivative in (\ref{moment-of-inertia-2})
can be interpreted as the derivative of the effective action
with respect to the source to obtain the electric
isovector form factor (at $q^2=0$) whereas the second
$\epsilon$ derivative is analogous to the expansion in angular
velocity. A detailed calculation including the valence quark contribution
gives:
\begin{equation}
\Theta\delta_{ab} = \frac{N_c}{2}
   \biggl\{ \sum\limits_{m \neq n}
\langle m |\tau^a| n \rangle
\langle n |\tau^b| m \rangle
 {\cal R}_\Theta^\Lambda(\varepsilon_m,\varepsilon_n)
+ \sum\limits_{n\neq val}
\frac{1}{\varepsilon_n - \varepsilon_{val}}
\langle n |\tau^a| val \rangle
\langle val |\tau^b| n \rangle \,.
\biggr\}
\label{moment-of-inertia-result}
\end{equation}
This regularized expression has been already derived in
ref.~\cite{Reinhardt89}
and numerically evaluated in refs.~\cite{Goeke91,Wakamatsu91}.

One can see from (\ref{electric-isovector-form-factor}) and
(\ref{moment-of-inertia-result}) that the isovector electric form
factor has a proper normalization :
\begin{equation}
G_E^{T=1}(0) = 1
\end{equation}
as it should be.

Now we proceed to compute the magnetic form factors.
 From matrix element (\ref{isoscalar-form-factor-2}) and
the definition of the magnetic Sachs form factor
(\ref{Sachs-magnetic-form-factor}) we find that the first non-vanishing
contributions come from rotational corrections:
\begin{equation}
G_M^{T=0}(q^2) = \frac{N_c{\cal M}_N}{6\Theta} \, \epsilon^{kaj} \,
\frac{iq^j}{|q^2|}
\int \d^3x \e^{i{\vec q}{\vec x}}
\sum\limits_{\scriptstyle \varepsilon_n > \varepsilon_{val}
\atop \varepsilon_m \leq \varepsilon_{val}   }
\frac{\Phi_m^\dagger({\vec x})\gamma^0\gamma^k \Phi_n({\vec x})
\, \langle n |\tau^a | m \rangle}
{\varepsilon_m - \varepsilon_n}   \,.
\label{magnetic-isoscalar-form-factor}
\end{equation}
The calculations of the magnetic isovector form factor are more
involved and the final expression includes leading order terms as well
as rotational corrections (next to leading order):
\begin{eqnarray}
&&G_M^{T=1}(q^2) = \frac{N_c{\cal M}_N}{3} \, \epsilon^{kbj} \,
\frac{iq^j}{|q^2|}
\int \d^3x \e^{i{\vec q}{\vec x}}
\Biggl\{
[\Phi_{val}^\dagger({\vec x})\gamma^0\gamma^k\tau^b \Phi_{val}({\vec x})]
 - \sum\limits_n {\cal R}_{M1}^\Lambda(\varepsilon_n) \,
[\Phi_n^\dagger({\vec x})\gamma^0\gamma^k\tau^b  \Phi_n({\vec x})]
\nonumber \\
&&-\frac{i}{2\Theta} \epsilon_{abc}\Biggl[ \sum\limits_{n,m}
{\cal R}_{M2}^\Lambda(\varepsilon_m,\varepsilon_n) \,
[\Phi_m^\dagger({\vec x})\gamma^0\gamma^k \tau^a \Phi_n({\vec x})]
\,\langle n |\tau^c |m\rangle
-\sum\limits_{\varepsilon_n \neq \varepsilon_{val}}
\frac{\mbox{sign}(\varepsilon_n) }
{\varepsilon_{val} - \varepsilon_n} \,
[\Phi_n^\dagger({\vec x})\gamma^0\gamma^k \tau^a \Phi_{val}({\vec x})]
\,\langle val |\tau^c | n \rangle
\Biggr]\Biggr\} \,.
\label{isovector-magnetic-form-factor}
\end{eqnarray}
Here, two different regularization functions appear:
\begin{equation}
{\cal R}_{M1}^\Lambda(\varepsilon_n) = - {\cal R}_1(\varepsilon_n,1)
= \varepsilon_n
\int\limits_{-\infty}^{\infty}\frac{\d\omega}{2\pi} \>
\frac{\e^{-(\omega^2+\varepsilon_n^2)/\Lambda^2}}{\omega^2+\varepsilon_n^2}
=\frac{\varepsilon_n}{2\sqrt{\pi}}
\int\limits_{1/\Lambda^2}^{\infty}\frac{\d u}{\sqrt{u}}
\e^{-u\varepsilon_n^2}
\,,
\end{equation}
\begin{eqnarray}
{\cal R}_{M2}^\Lambda(\varepsilon_m,\varepsilon_n) &=& \frac{1}{4}
\left[- {\cal R}_2^{(+)}(\varepsilon_m,\varepsilon_n,1)
 + {\cal R}_2^{(-)}(\varepsilon_m,\varepsilon_n,1)\right]
\nonumber\\
&=& \frac{1}{4\pi} \int_0^1 \frac{\d\beta}{\sqrt{\beta(1-\beta)}} \,
\frac{(1-\beta)\varepsilon_m - \beta\varepsilon_n}
{(1-\beta)\varepsilon_m^2 + \beta\varepsilon_n^2} \,
\e^{-[(1-\beta)\varepsilon_m^2 + \beta\varepsilon_n^2]/\Lambda^2} \,.
\end{eqnarray}

Eqs.
(\ref{form-factor-electric-isosinglet}),
(\ref{electric-isovector-form-factor}),
(\ref{magnetic-isoscalar-form-factor}) and
(\ref{isovector-magnetic-form-factor})
are our final expressions for the electromagnetic
isoscalar and isovector form factors in the NJL model.
According to (\ref{isoscalar-isovector})
the proton and neutron form factors are expressed in terms of the isoscalar
and isovector form factors as follows
\begin{equation}
G_{E(M)}^p = \frac12 [G_{E(M)}^{T=0} +  G_{E(M)}^{T=1}] \,,
\end{equation}
\begin{equation}
G_{E(M)}^n = \frac12 [G_{E(M)}^{T=0} -  G_{E(M)}^{T=1}]\,.
\label{FFPNb}
\end{equation}

\section{Numerical results}

In the numerical calculations we use the method of Ripka and Kahana
{}~\cite{Ripka84} for solving the eigenvalue problem in finite quasi--discrete
basis. We consider a spherical box of large radius
$D$ and the basis is made discrete by imposing a boundary condition at
$r=D$. Also, it is made finite by restricting momenta of the basis states
to be smaller than the numerical cut-off $K_{max}$.
Both quantities have no physical meaning and the results
should not depend on them. The typical values used are
$D \sim 20/M$ and $K_{max} \sim 7 M$.

In addition, all checks concerning the numerical stability of the
solution with respect to varying box size and choice of the
numerical cut-off have been done and the actual calculation is
completely under control.

The parameters of the model are fixed in meson sector in the well known
way ~\cite{TMeissner89} to have $f_{\pi} = 93$ MeV and $m_{\pi} =
139.6$ MeV. This leaves the constituent quark mass $M$ as the only free
parameter.

The proton and neutron electric and magnetic form factors
are displayed in Figs.\ref{Figr2}-\ref{Figr5}. The theoretical curves
resulting from the model are given for four different values of the
constituent quark mass, namely $370, 400, 420$ and $450$ MeV.
The magnetic form factors are normalized to the experimental
values of the corresponding magnetic moments at $q^2=0$.
With one exception of the neutron electric form factor (Fig.\ref{Figr3}),
all other form factors agree with the experimental data quite well. The best
fit is for the constituent quark mass around $420$ MeV.

As can be seen the only form factor which deviates from the experimental
data is the neutron electric form factor and this requires
some explanation. Obviously, this form factor is the most sensitive
for numerical errors.  According to the formula (\ref{FFPNb})
the form factor has been calculated as a difference of the electric
isoscalar and electric isovector form factors. Both form factors were of
order of one and calculated by the code with high enough accuracy.
However, the resulting neutron form factor has experimental
values of order $0.04$, {\it i.e.} about $4\%$ of the value of its
components. This means that even small theoretical uncertainties for one of
the components can be enhanced by a factor $50$.
It means that the numerical accuracy together with the used
large $1/N_c$ approximation behind the model
are strongly magnified and could in principle result in a deviation from
the experimental data
for momentum transfers above $100$ MeV. It should be also stressed that the
experimental data (see ref.~\cite{Platchkov90}), available for
the neutron electric form factor, are strongly model
dependent and a different N - N potential used
in the analysis of the data can lead to an enhancement of the experimental
numbers by more than 50 \%. There are also very recent
experiment~\cite{Eden94}, which deviate from the previous
data~\cite{Platchkov90}, indicating larger values closer to our numbers.

As the next step, we compute other electromagnetic observables:
the mean squared radii, the magnetic moments and the
nucleon--$\Delta$ splitting. In particular, the charge radii and the
magnetic moments can be obtained from the form factors:
\begin{equation}
\langle r^2 \rangle_{T=0,1} = -{6\over G_E^{T=0,1}}\, {\d G_E^{T=0,1}\over
\d q^2}\, \Bigg\vert_{q^2=0} \ ,
\label{ELRAD} \end{equation}
\begin{equation}
\mu^{T=0,1} = G_M^{T=0,1} (q^2) \Big\vert_{q^2=0} \ .
\label{MAGMOM} \end{equation}

For the quark masses $370, 420$ and $450$
MeV and pion mass $m_{\pi}=140$ MeV the calculated values are
presented in Table~\ref{Tabl1}. The values for the axial vector coupling
constant $g_A$, taken from ref.~\cite{Christov94}, are also  presented
for completeness.

The results of Table~\ref{Tabl1}   ($m_{\pi}=139.6$ MeV) again indicate the
value $\sim 420$ MeV for the constituent quark mass, in
agreement with the conclusion drawn from the form factor curves.
The same value has been suggested earlier ~\cite{Goeke91,Gorski92},
where a smaller number of observables has been considered.
With the exception of the neutron electric squared radii, to
which remarks similar to the case of the neutron electric form factor
are valid, the contribution of the valence quarks is dominant. However,
the contribution of the Dirac sea is non-negligible and it varies within
the range 15 -- 40\%. As can be seen, the numerical results for
the nucleon $N$--$\Delta$ mass splitting ($M_{\Delta}-M_N$),
the mean squared proton, isoscalar and isovector
electric radii and the axial coupling constant ($g_A$) as well as the
$q$-dependence of the proton electric and magnetic, and neutron
magnetic form factor differ from the experimental data by no more then
about $\pm 5\%$. Finally, for the magnetic moments we have got results
20--25\% below their experimental values. However, despite
of this underestimation of both magnetic moments, we have found a very good
result for the ratio $\mu_p/\mu_n$ which is far better than in other models.
The experimental ratio is almost exactly reproduced for the constituent
quark mass 420 MeV.

In the present calculations, in
contrast to the isoscalar magnetic moment the isovector one includes
non-zero contributions in both leading ($N_c^0$) and next to leading order
($1/N_c$ rotational corrections). The enhancement due to these corrections
improves considerably (see also ref.~\cite{Christov94})
the agreement with the experiment and resolves to a great
extent the problem of strong underestimation~\cite{Wakamatsu91} of both the
isovector magnetic moment and the axial vector coupling constant in leading
order. It is also important to note that since the $1/N_c$ rotational
corrections have the same spin-flavor structure like the leading term
they do not violate the consistency condition of Dashen and
Manohar~\cite{Dashen93} derived in the large-$N_c$ of QCD. The latter means
that other $1/N_c$ corrections (e.g. meson loops) should contribute to both
the leading and next to leading order.

In Table~\ref{Tabl2} we give the theoretical values of the same
quantities but with the physical pion mass set to zero.
The chiral limit ($m_{\pi}\to 0$) mostly
influences the isovector charge radius. In fact, as it should be
expected~\cite{Bet72,Adkins83}, the isovector charge radius diverges in
chiral limit and our calculations with zero pion mass confirms it.
It can be seen in Fig.\ref{Figr6},
where the isovector charge radius is plotted {\it vs.} the box size $D$.
The Dirac sea contribution to the radius grows linearly with $D$ and
diverges as $D\to\infty$. Because of this that quantity (and the relative
quantities) is not included in Table~\ref{Tabl2}.
The other observable strongly influenced by the chiral limit
is the neutron electric form factor.
For the $m_{\pi}\to 0$ the discrepancy from the experiment is
by almost a factor two larger than in the case $m_{\pi}\neq 0$.
The other observables differ in the chiral limit by about 30\%.
The comparison of the values in the two tables indicate that
taking the physical pion mass gives us a best fit with
a much better agreement with the experimental data. In
addition, in chiral limit we observe much larger contribution from the sea
effects, about 50\% of the total value.

In section 3 we have found that the magnetic isovector form factor is the
only one which includes non-zero contributions in both leading ($N_c^0$) and
next to leading order ($1/N_c$) rotational corrections. However, the
numerical
calculations show that the ($1/N_c$) rotational corrections do not
affect the $q^2$-dependence (slope) of the form factor but rather the
value at the origin, $G_M^{T=1}(0)\,=\,\mu^{T=1}$ which is the
isovector magnetic
moment. It is not surprising since (as can be seen from
eq. (\ref{isovector-magnetic-form-factor})
in both leading and next to leading order terms the shape of the wave
functions $\Phi_n$ determines the $q^2$-dependence of the form factor
whereas the value at $q^2=0$ depends on the particular matrix elements
included. In leading order the isovector magnetic moment is
strongly underestimated~\cite{Wakamatsu91}. As has been shown in
ref.~\cite{Christov94} the enhancement for this quantity
due to the $1/N_c$ corrections is of order $(N_c+2)/N_c$ and it improves
considerably the agreement with experiment. However, as can be seen from
Table \ref{Tabl1} the magnetic moments are still below the experimental
value by 25 \%.

The isoscalar and isovector electric mean squared radii are
shown in Figs.\ref{Figr7},\ref{Figr8} as functions of the constituent
quark mass. The same plot for the proton and neutron electric charge
radii is given in Fig.\ref{Figr9}.
In these plots the valence and sea contributions are explicitly
given (dashed and dash-dotted lines, respectively). As could be
expected,
for the isoscalar electric charge radius the valence part is dominant
(about 85\%), due to the fact that there is no $1/N_c$ rotational
contributions. This is not true for the isovector electric charge
radius, where the sea part contributes to about 45\% of the total value
(Fig.\ref{Figr7}). Also, this effect can be seen at the proton and neutron
charge radii (Fig.\ref{Figr9}) which are linear combinations of the
isoscalar and isovector ones.
For proton the sea contribution is about 30\% whereas for the
neutron charge radius the negative sea part is dominating and the valence
contribution is negligible.

In addition, in Fig.\ref{Figr10} we plot the proton and neutron charge
distribution for the constituent quark  mass $M\,=\,420\,$ MeV. For the
proton we have a positive definite charge distribution completely dominated
by the valence contribution whereas in the case of the neutron the Dirac sea
is dominant. In accordance with the well accepted phenomenological
picture one realizes a positive core coming
from the valence quarks and a long negative tail due to the polarization
of the Dirac sea. Using the gradient expansion the latter can be expressed
in terms of the dynamical pion field -- pion cloud.

For completeness, in Fig.\ref{Figr11} we present also the magnetic moment
density distribution for proton and neutron for the constituent quark
mass $M\,=\,420\,$ MeV. The sea contribution becomes non-negligible for
distances greater than 0.5 fm. Due to the relatively large tail
contribution to the magnetic moments the sea contribution to these quantities
is about 30\%.

\section{Summary and conclusions}

Our numerical results support the view that the chiral quark soliton model
of the SU(2) Nambu--Jona-Lasinio type offers relatively simple but
quite successful description of some low-energy QCD phenomena and,
in particular, of the nucleon electromagnetic properties.
In calculations, using $f_{\pi} =
93$ MeV, $m_{\pi} = 139.6$ MeV and a constituent quark mass of $M = 420$
MeV, and no other parameters, we have obtained an overall good description
for the
electromagnetic form factors, the mean squared radii, the magnetic moments,
and the nucleon--$\Delta$ splitting. The isoscalar and isovector
electric radii are reproduced within 15\%. The magnetic moments are
underestimated by about 25\%. However, their ratio
$\mu_p/\mu_n$ is almost perfectly reproduced. To this end, the account of
the $1/N_c$ rotational corrections is decisive. The $q$--dependence
of $G^p_E(q^2)$, $G^p_M(q^2)$ and $G^n_M(q^2)$ is very well reproduced up
to momentum transfer of 1 GeV. The only exception is the neutron form
factor $G^n_E(q^2)$ which is by a
factor of two too large for $q^2 > 100$ MeV$^2$. One should note, however,
that $G^n_E(q^2)$ is more than an order of magnitude smaller
than $G^p_E(q^2)$  and as such it is extremely sensitive
to both the model approximations and the numerical methods used.
Here, it turns out that the agreement
is noticeably worse if the chiral limit ($m_{\pi}~\to~0$) is
used. This can be easily understood since the pion mass
determines the asymptotic behavior of the pion field.
Altogether we conclude that the
chiral quark soliton model based on a bosonized SU(2) NJL type lagrangean is
quite appropriate for the evaluation of nucleonic electromagnetic
properties.

\noindent{\bf Acknowledgement:}

The project has been partially supported by the BMFT, DFG and COSY
(J\"ulich). Also, we are greatly indebted for financial support to
the Bulgarian Science Foundation $\Phi$--32
(CVC) and to the Polish Committee for
Scientific Research,  Projects KBN nr 2 P302 157 04
and 2 0091 91 01 (AZG).

\newpage
\begin{figure}
\caption{Diagrams corresponding to the expansion in $\Omega$ of the current
matrix element:
a) the valence contribution and b) the Dirac sea contribution.}
\label{Figr1}
\end{figure}
\begin{figure}
\caption{The proton electric form factor for the
momentum transfers below 1 GeV. Experimental data from
ref.~\protect{\cite{Hoehler76}}.} \label{Figr2}
\end{figure}
\begin{figure}
\caption{ The neutron electric form factor for the
momentum transfers below 1 GeV. experimental data from
refs.~\protect{\cite{Platchkov90,Eden94}}.} \label{Figr3}
\end{figure}
\begin{figure}
\caption{ The proton magnetic form factor normalized to the experimental
value of the proton magnetic moment at $q^2=0$ for the momentum transfers
below 1 GeV. The normalization factor can be extracted from Table
1. Experimental data from ref.~\protect{\cite{Hoehler76}}.} \label{Figr4}
\end{figure}
\begin{figure}
\caption{ The neutron magnetic form factor normalized to the experimental
value of the neutron magnetic moment at $q^2=0$ for the momentum transfers
below 1 GeV. The normalization factor can be extracted from Table 1.
Experimental data from ref.~\protect{\cite{Hoehler76}}.} \label{Figr5}
\end{figure}
\begin{figure}
\caption{ The isovector charge radius as a function of
the box size $D$ for $m_{\pi}=0$ and $m_{\pi}=139.6$ MeV.}
\label{Figr6}
\end{figure}
\begin{figure}
\caption{ The isoscalar electric charge radius as a function of
the constituent quark mass $M$. The valence and sea parts are marked
by the dashed and dashed-dotted lines.}
\label{Figr7}
\end{figure}
\begin{figure}
\caption{ The isovector electric charge radius as a function of
the constituent quark mass $M$. The valence and sea parts are marked
by the dashed and dashed-dotted lines.}
\label{Figr8}
\end{figure}
\begin{figure}
\caption{ The electric charge radii of proton and neutron
as functions of the constituent quark mass $M$. The valence and sea
parts are marked by the dashed lines.}
\label{Figr9}
\end{figure}
\begin{figure}
\caption{ The charge density distribution of the proton (lower) and
neutron (upper) for the constituent quark mass $M = 420$ MeV.}
\label{Figr10}
\end{figure}
\begin{figure}
\caption{ The magnetic moment density of proton and neutron
for the constituent quark mass $M = 420$ MeV.}
\label{Figr11}
\end{figure}

\newpage
\begin{table}
\caption{ Nucleon observables calculated with the physical
pion mass.}
\vbox{\offinterlineskip
\hrule height1pt
\halign{&\vrule width1pt#&
 \strut\quad\hfill#\hfill\quad
 &\vrule#&\quad\hfill#\hfil\quad
 &\vrule#&\quad\hfill#\hfil\quad
 &\vrule#&\quad\hfill#\hfil\quad
 &\vrule#&\quad\hfill#\hfil\quad
 &\vrule#&\quad\hfill#\hfil\quad
 &\vrule#&\quad\hfill#\hfil\quad
 &\vrule width1pt#&\quad\hfil#\hfil\quad\cr
height6pt&\omit&&\multispan{11}&&\omit&\cr
& &&\multispan{11} \hfill {\bf Constituent\ Quark\ Mass} \hfill &&
&\cr
height4pt&\omit&&\multispan{11}&&\omit&\cr
&\hfill {\bf Quantity}\hfill &&\multispan3 \hfill 370\ MeV\hfill
&&\multispan3
\hfill 420   MeV\hfill
&&\multispan3
\hfill 450  MeV\hfill
&&\hfill {\bf Exper.}\hfill &\cr
height3pt&\omit&&\multispan3 && \multispan3 &&
\multispan3 &&\omit&\cr
&\omit&&\multispan{11}\hrulefill&&\omit&\cr
height3pt&\omit&&\omit&&\omit&&\omit&&\omit&&\omit&&\omit&&\omit&\cr
&\hfill    &&\hfill total \hfill &&\hfill sea \hfill &&\hfill total
\hfill &&\hfill sea \hfill && \hfill total \hfill  &&\hfill sea \hfill  &&
&\cr %
height3pt&\omit&&\omit&&\omit&&\omit&&\omit&&\omit&&\omit&&\omit&\cr
\noalign{\hrule height1pt}
height3pt&\omit&&\omit&&\omit&&\omit&&\omit&&\omit&&\omit&&\omit&\cr

& \hfill $ <r^2>_{T=0}$ [fm$^2$]  &&
0.63  &&
0.05  &&
0.52  &&
0.07  &&
0.48  &&
0.09  &&
0.62  &\cr
height3pt&\omit&&\omit&&\omit&&\omit&&\omit&&\omit&&\omit&&\omit&\cr
\noalign{\hrule}
height3pt&\omit&&\omit&&\omit&&\omit&&\omit&&\omit&&\omit&&\omit&\cr

& \hfill $ <r^2>_{T=1}$ [fm$^2$]  &&
1.07  &&
0.33  &&
0.89  &&
0.41  &&
0.84  &&
0.45  &&
0.86 &\cr
height3pt&\omit&&\omit&&\omit&&\omit&&\omit&&\omit&&\omit&&\omit&\cr
\noalign{\hrule}
height3pt&\omit&&\omit&&\omit&&\omit&&\omit&&\omit&&\omit&&\omit&\cr

& \hfill $ <r^2>_p$ [fm$^2$]  &&
 0.85  &&
 0.19  &&
 0.70  &&
 0.24  &&
 0.66  &&
 0.27  &&
 0.74  &\cr
height3pt&\omit&&\omit&&\omit&&\omit&&\omit&&\omit&&\omit&&\omit&\cr
\noalign{\hrule}
height3pt&\omit&&\omit&&\omit&&\omit&&\omit&&\omit&&\omit&&\omit&\cr

& \hfill $ <r^2>_n$ [fm$^2$]  &&
--0.22   &&
--0.14   &&
--0.18   &&
--0.17   &&
--0.18   &&
--0.18   &&
--0.12   &\cr
height3pt&\omit&&\omit&&\omit&&\omit&&\omit&&\omit&&\omit&&\omit&\cr
\noalign{\hrule}
height3pt&\omit&&\omit&&\omit&&\omit&&\omit&&\omit&&\omit&&\omit&\cr

& \hfill $\mu_{T=0}\hfill$ [n.m.]  &&
0.68  &&
0.09  &&
0.62  &&
0.03  &&
0.59  &&
0.05  &&
0.88  &\cr
height3pt&\omit&&\omit&&\omit&&\omit&&\omit&&\omit&&\omit&&\omit&\cr
\noalign{\hrule}
height3pt&\omit&&\omit&&\omit&&\omit&&\omit&&\omit&&\omit&&\omit&\cr

& \hfill $ \mu_{T=1}\hfill$ [n.m.]  &&
 3.56 &&
 0.77 &&
 3.44 &&
 0.97 &&
 3.16 &&
 0.80 &&
 4.71        &\cr
height3pt&\omit&&\omit&&\omit&&\omit&&\omit&&\omit&&\omit&&\omit&\cr
\noalign{\hrule}
height3pt&\omit&&\omit&&\omit&&\omit&&\omit&&\omit&&\omit&&\omit&\cr

& \hfill $ \mu_p \hfill$ [n.m.]  &&
  2.12 &&
  0.43 &&
  2.03 &&
  0.50 &&
  1.86 &&
  0.43 &&
  2.79        &\cr
height3pt&\omit&&\omit&&\omit&&\omit&&\omit&&\omit&&\omit&&\omit&\cr
\noalign{\hrule}
height3pt&\omit&&\omit&&\omit&&\omit&&\omit&&\omit&&\omit&&\omit&\cr
& \hfill $ \mu_n \hfill$ [n.m.]  &&
 --1.44 &&
 --0.34 &&
 --1.41 &&
 --0.47 &&
 --1.29 &&
 --0.38 &&
 --1.91    &\cr
height3pt&\omit&&\omit&&\omit&&\omit&&\omit&&\omit&&\omit&&\omit&\cr
\noalign{\hrule}
height3pt&\omit&&\omit&&\omit&&\omit&&\omit&&\omit&&\omit&&\omit&\cr
& \hfill $ |\mu_p/\mu_n|$ \hfill  &&
  1.47  &&
  ---   &&
  1.44  &&
  ---   &&
  1.44  &&
  ---   &&
  1.46           &\cr
height3pt&\omit&&\omit&&\omit&&\omit&&\omit&&\omit&&\omit&&\omit&\cr
\noalign{\hrule}
height3pt&\omit&&\omit&&\omit&&\omit&&\omit&&\omit&&\omit&&\omit&\cr
& \hfill $ <r^2>^\mu_p$ [fm$^2$]  &&
 1.08  &&
 0.32  &&
 0.66  &&
 0.28  &&
 0.56  &&
 0.25  &&
 0.74  &\cr
height3pt&\omit&&\omit&&\omit&&\omit&&\omit&&\omit&&\omit&&\omit&\cr
\noalign{\hrule}
height3pt&\omit&&\omit&&\omit&&\omit&&\omit&&\omit&&\omit&&\omit&\cr

& \hfill $ <r^2>^\mu_n$ [fm$^2$]  &&
1.17   &&
0.51   &&
0.65   &&
--0.31   &&
--0.52   &&
--0.24   &&
--0.77   &\cr
height3pt&\omit&&\omit&&\omit&&\omit&&\omit&&\omit&&\omit&&\omit&\cr
\noalign{\hrule}
height3pt&\omit&&\omit&&\omit&&\omit&&\omit&&\omit&&\omit&&\omit&\cr

& \hfill $ M_{\Delta}-M_N$ [MeV] &&
 213  &&
 ---  &&
 280  &&
 ---  &&
 314  &&
 ---  &&
 294  &\cr
height3pt&\omit&&\omit&&\omit&&\omit&&\omit&&\omit&&\omit&&\omit&\cr
\noalign{\hrule}
height3pt&\omit&&\omit&&\omit&&\omit&&\omit&&\omit&&\omit&&\omit&\cr

& \hfill $ g_A \hfill $ &&
 1.26  &&
 0.08  &&
 1.21  &&
 0.11  &&
 1.13  &&
 0.06  &&
 1.26         &\cr
height3pt&\omit&&\omit&&\omit&&\omit&&\omit&&\omit&&\omit&&\omit&\cr
\noalign{\hrule height1pt} }}
\label{Tabl1}
\end{table}

\begin{table}
\caption{ Nucleon observables calculated with the zero
pion mass.}
\vbox{\offinterlineskip
\hrule height1pt
\halign{&\vrule width1pt#&
 \strut\quad\hfil#\quad
 &\vrule#&\quad\hfil#\quad
 &\vrule#&\quad\hfil#\quad
 &\vrule#&\quad\hfil#\quad
 &\vrule#&\quad\hfil#\quad
 &\vrule#&\quad\hfil#\quad
 &\vrule#&\quad\hfil#\quad
 &\vrule#&\quad\hfil#\quad
 &\vrule width1pt#&\quad\hfil#\quad\cr
height6pt&\omit&&\multispan{11}&&\omit&\cr
& &&\multispan{11} \hfill {\bf Constituent\ Quark\ Mass} \hfill &&
&\cr
height4pt&\omit&&\multispan{11}&&\omit&\cr
&\hfill {\bf Quantity}\hfill &&\multispan3 \hfill 370\ MeV\hfill
&&\multispan3
\hfill 420   MeV\hfill
&&\multispan3
\hfill 450   MeV\hfill
&&\hfill {\bf Exper.} &\cr
height3pt&\omit&&\multispan3 && \multispan3 &&
\multispan3 &&\omit&\cr
&\omit&&\multispan{11}\hrulefill&&\omit&\cr
height3pt&\omit&&\omit&&\omit&&\omit&&\omit&&\omit&&\omit&&\omit&\cr
&\hfill    &&\hfill total &&\hfill sea &&\hfill total &&\hfill sea &&
\hfill total &&\hfill sea &&   &\cr
height3pt&\omit&&\omit&&\omit&&\omit&&\omit&&\omit&&\omit&&\omit&\cr
\noalign{\hrule height1pt}
height3pt&\omit&&\omit&&\omit&&\omit&&\omit&&\omit&&\omit&&\omit&\cr

& \hfill $ <r^2>_{T=0}$ \hfill [fm$^2$]  &&
\hfill 0.88   &&
\hfill 0.20  &&
\hfill 0.66  &&
\hfill 0.26  &&
\hfill 0.61  &&
\hfill 0.23  &&
\hfill 0.62\hfill&\cr
height3pt&\omit&&\omit&&\omit&&\omit&&\omit&&\omit&&\omit&&\omit&\cr
\noalign{\hrule}
height3pt&\omit&&\omit&&\omit&&\omit&&\omit&&\omit&&\omit&&\omit&\cr

& \hfill $ \mu_{T=0}$ \hfill [n.m.]  &&
\hfill 0.66  &&
\hfill 0.07  &&
\hfill 0.59  &&
\hfill 0.09  &&
\hfill 0.57  &&
\hfill 0.09  &&
\hfill 0.88\hfill
&\cr
height3pt&\omit&&\omit&&\omit&&\omit&&\omit&&\omit&&\omit&&\omit&\cr
\noalign{\hrule}
height3pt&\omit&&\omit&&\omit&&\omit&&\omit&&\omit&&\omit&&\omit&\cr

& \hfill $ \mu_{T=1}$ \hfill [n.m.]  &&
\hfill 4.61   &&
\hfill 1.59  &&
\hfill 4.38  &&
\hfill 1.89  &&
\hfill 3.91  &&
\hfill 1.48  &&
\hfill 4.71 \hfill&\cr
height3pt&\omit&&\omit&&\omit&&\omit&&\omit&&\omit&&\omit&&\omit&\cr
\noalign{\hrule}
height3pt&\omit&&\omit&&\omit&&\omit&&\omit&&\omit&&\omit&&\omit&\cr

& \hfill $ \mu_p$ \hfill [n.m.]  &&
\hfill   2.63  &&
\hfill   0.83  &&
\hfill   2.49  &&
\hfill   0.99  &&
\hfill   2.24  &&
\hfill   0.79  &&
\hfill   2.79 \hfill &\cr
height3pt&\omit&&\omit&&\omit&&\omit&&\omit&&\omit&&\omit&&\omit&\cr
\noalign{\hrule}
height3pt&\omit&&\omit&&\omit&&\omit&&\omit&&\omit&&\omit&&\omit&\cr

& \hfill $ \mu_n$ \hfill [n.m.]  &&
\hfill --1.97  &&
\hfill --0.76  &&
\hfill --1.89  &&
\hfill --0.90  &&
\hfill --1.67  &&
\hfill --0.70  &&
\hfill --1.91 \hfill &\cr
height3pt&\omit&&\omit&&\omit&&\omit&&\omit&&\omit&&\omit&&\omit&\cr
\noalign{\hrule}
height3pt&\omit&&\omit&&\omit&&\omit&&\omit&&\omit&&\omit&&\omit&\cr

& \hfill $ |\mu_p/\mu_n|$ \hfill  &&
\hfill 1.34 &&
\hfill ---  &&
\hfill 1.34  &&
\hfill ---  &&
\hfill 1.34  &&
\hfill ---  &&
\hfill 1.46 \hfill &\cr
height3pt&\omit&&\omit&&\omit&&\omit&&\omit&&\omit&&\omit&&\omit&\cr
\noalign{\hrule}
height3pt&\omit&&\omit&&\omit&&\omit&&\omit&&\omit&&\omit&&\omit&\cr

& \hfill $ M_{\Delta}-M_N$ \hfill [MeV]  &&
\hfill 221\hfill &&
\hfill ---\hfill &&
\hfill 261\hfill &&
\hfill ---\hfill &&
\hfill 301\hfill &&
\hfill ---\hfill &&
\hfill 294\hfill &\cr
height3pt&\omit&&\omit&&\omit&&\omit&&\omit&&\omit&&\omit&&\omit&\cr
\noalign{\hrule}
\noalign{\hrule height1pt} }}
\label{Tabl2}
\end{table}


\begin{thebibliography}{99}


\bibitem{Nambu61}Y.Nambu and G.Jona-Lasinio, {\it Phys.Rev.} {\bf 122}
(1961) 354

\bibitem{Eguchi74} T.Eguchi, H.Sugawara,
{\it Phys.Rev.} {\bf D10} (1974) 4257;
T.Eguchi, {\it Phys.Rev.} {\bf D14} (1976) 2755

\bibitem{Diakonov84} D.Diakonov, V.Petrov,
{\it Phys.Lett.} {\bf 147B} (1984) 351; {\it Nucl.Phys.} {\bf B272} (1986)
457; Quark Cluster Dynamics, eds. K.Goeke, P.Kroll,H.-R.Petry,
Lecture Notes in Physics, Springer-Verlag, v.417, 1992, p.288

\bibitem{Diakonov89}D.I.Diakonov, V.Yu.Petrov and P.V.Pobylitsa, 21$^{st}$
LNPI Winter School on Elem.Part. Physics, Leningrad 1986; {\it Nucl.Phys.}
{\bf B306}(1988)809;

\bibitem{Reinhardt88}H.Reinhardt and R.W\"unsch, {\it Phys.Lett.} {\bf
215B}(1988)577; ibid. {\bf B230} (1989) 93

\bibitem{TMeissner89} T.Meissner, F.Gr\"ummer and K.Goeke, Phys.Lett.
{\bf 227B} (1989) 296; {\it Ann.Phys.} {\bf 202} (1990) 297; T.Meissner and
K.Goeke, Nucl.Phys. {\bf A254} (1991) 719

\bibitem{Reinhardt89}H.Reinhardt, {\it Nucl.Phys.} {\bf
503B}(1989)825

\bibitem{Goeke91} K.Goeke, A.Z.G\'orski, F.Gr\"ummer,
Th.Meissner, H.Reinhardt, R.W\"unsch,
{\it Phys.Lett.} {\bf B256} (1991) 321

\bibitem{TMeissner91} T.Meissner and K.Goeke, Z.Phys.{\bf A339}(1991)513

\bibitem{Gorski92} A.Z.G\'orski, K.Goeke, F.Gr\"ummer,
{\it Phys.Lett.} {\bf B278} (1992) 24

\bibitem{Christov94} Chr.V.Christov, A.Blotz, K.Goeke, V.Yu.Petrov,
P.V.Pobylitsa, M.Wakamatsu, T.Watabe, {\it Phys.Lett.} {\bf B325}(1994)467

\bibitem{Wakamatsu91} M.Wakamatsu, H.Yoshiki,
{\it Nucl.Phys.} {\bf A524} (1991) 561

\bibitem{Chu94} M.-C.Chu, J.M.Grandy, S.Huang and J.W.Negele, {\it
Phys.Rev.} {\bf D49} (1994) 6039

\bibitem{Ball89} R.Ball, {\it Phys.Rep.} {\bf 182} (1989) 1

\bibitem{Schaden90} M.Schaden, H.Reinhardt, P.Amundsen,
M.Lavelle, {\it Nucl.Phys.} {\bf B339} (1990) 595

\bibitem{Ripka84} S.Kahana and G.Ripka, Nucl.Phys.{\bf A429} (1984) 462

\bibitem{Hoehler76} G.H\"ohler at al., Nucl.Phys.{\bf B414} (1976) 505

\bibitem{Platchkov90} S.Platchkov at al., Nucl.Phys.{\bf A510} (1990) 740

\bibitem{Eden94} T.Eden at al., Phys.Rev.{\bf C50} (1994) R1749

\bibitem{Dashen93}R.Dashen and A.V.Manohar, {\it Phys.Lett.}{\bf 315B}
(1993) 425, 438

\bibitem{Bet72} M.A.Bet and A.Zepeda, {\it Phis.Rev.} {\bf D6} (1972) 2912

\bibitem{Adkins83} G.S.Adkins, C.R.Nappi and E.Witten, {\it Nucl.Phys.}
{\bf B228} (1983) 552


\end{thebibliography}
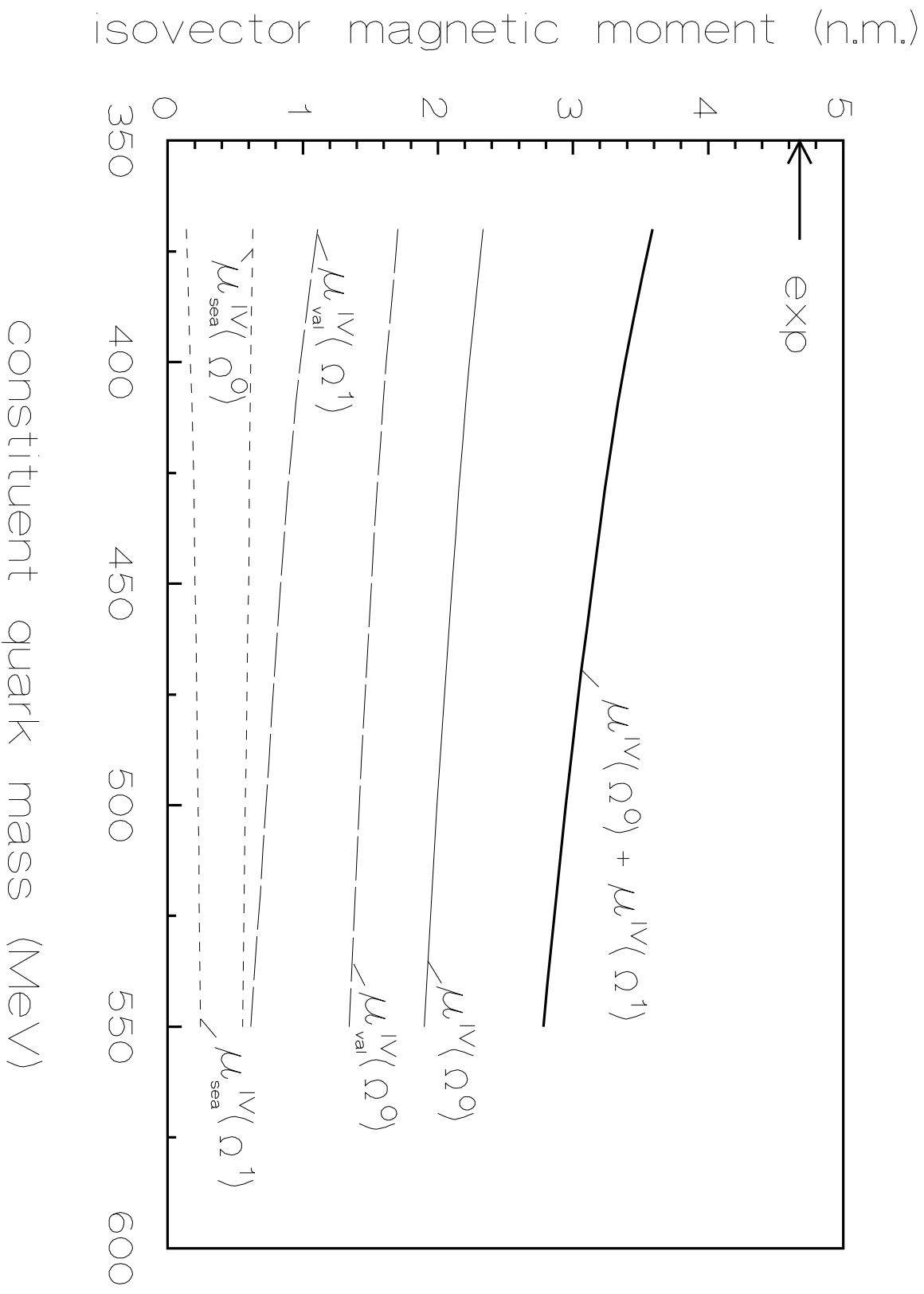
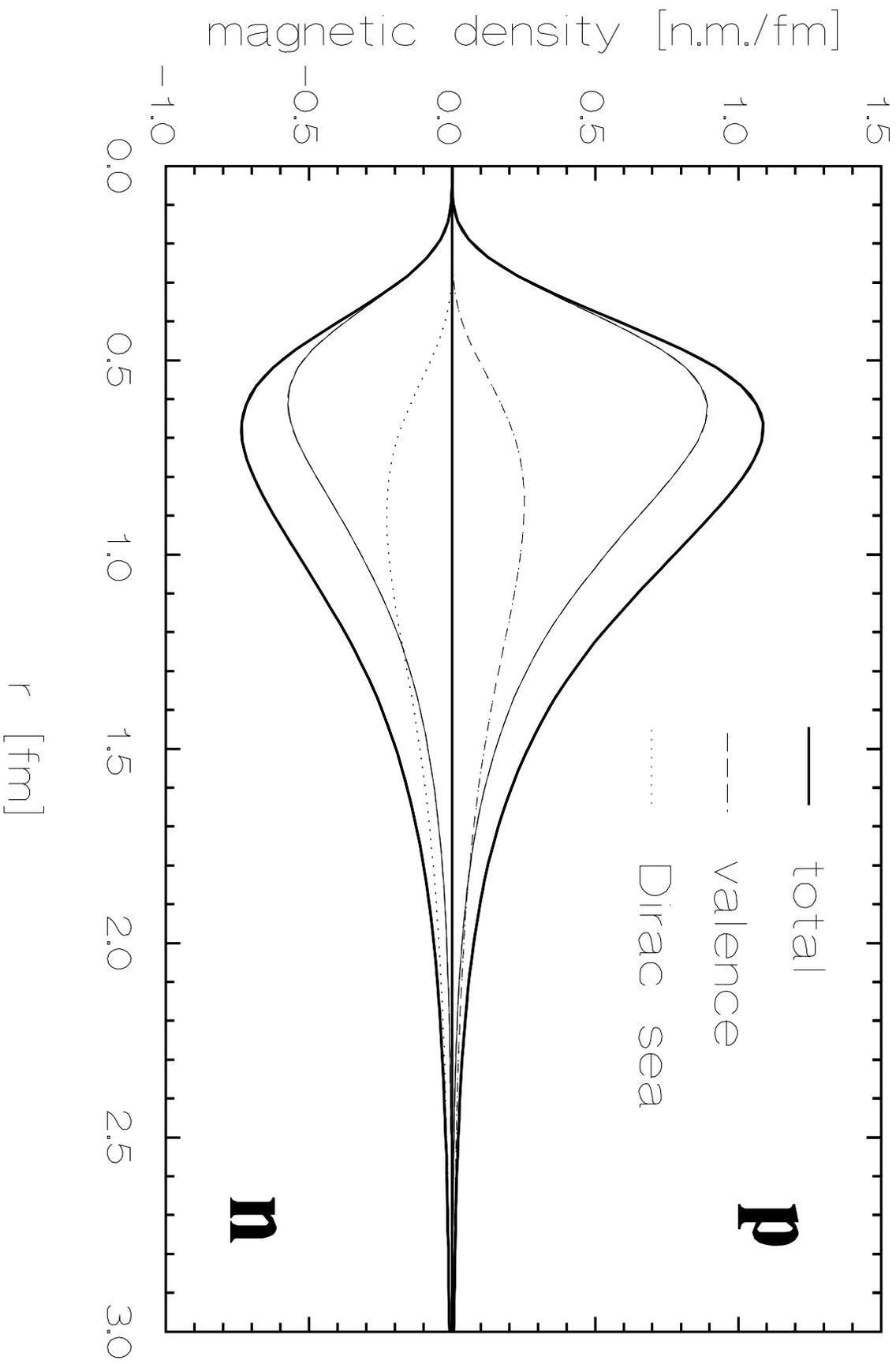
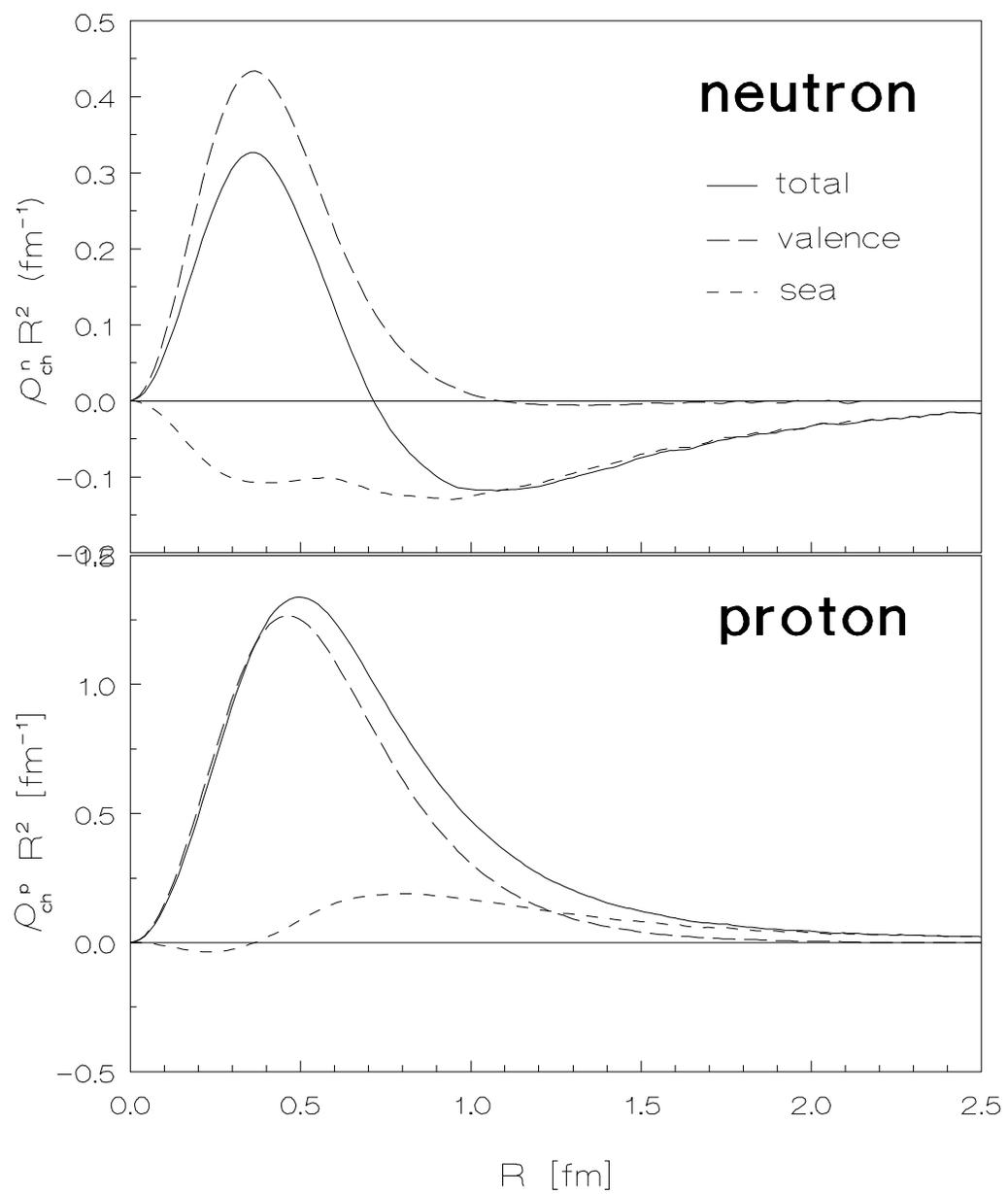
\end{document}